\documentclass[aps,pre,citeautoscript,twocolumn,showpacs,floatfix,superscriptaddress, graphics,float]{revtex4-1}

\usepackage{amsmath,amssymb,graphicx,color,braket,hyphenat,makeidx,xcolor,dsfont,subfigure}
\usepackage[ocgcolorlinks,colorlinks=true,linkcolor=blue,citecolor=red,linktocpage=true]{hyperref}

\newcommand{\R}{\mathcal{R}}
\newcommand{\T}{\mathcal{T}}
\newcommand{\gammat}{\gamma_{\{0,t\}}}
\newcommand{\gammatau}{\gamma_{\{0,\tau\}}}
\def\id {{\mathds 1}}

\begin{document}

\title{Thermodynamics of Gambling  Demons}

\author{Gonzalo Manzano}
\affiliation{International Centre for Theoretical Physics ICTP, Strada Costiera 11, I-34151, Trieste, Italy}
\affiliation{Institute for Quantum Optics and Quantum Information (IQOQI), Austrian Academy of Sciences, Boltzmanngasse 3, 1090 Vienna, Austria.}

\author{Diego Subero}
\affiliation{PICO group, QTF Centre of Excellence, Department of Applied Physics, Aalto University, 00076 Aalto, Finland}

\author{Olivier Maillet}
\affiliation{PICO group, QTF Centre of Excellence, Department of Applied Physics, Aalto University, 00076 Aalto, Finland}

\author{Rosario Fazio}
\affiliation{International Centre for Theoretical Physics ICTP, Strada Costiera 11, I-34151, Trieste, Italy}
\affiliation{Dipartimento di Fisica, Universit\`a di Napoli ``Federico II'', Monte S. Angelo, I-80126 Napoli, Italy}

\author{Jukka P. Pekola}
\affiliation{PICO group, QTF Centre of Excellence, Department of Applied Physics, Aalto University, 00076 Aalto, Finland}

\author{\'Edgar Rold\'an} 
\affiliation{International Centre for Theoretical Physics ICTP, Strada Costiera 11, I-34151, Trieste, Italy}

\begin{abstract}
We introduce and realize demons that follow a customary gambling strategy to stop a nonequilibrium process at stochastic times. We derive second-law-like inequalities for the average work done in the presence of gambling, and universal stopping-time fluctuation relations for classical and quantum non-stationary stochastic processes. We test experimentally our results in a single-electron box, where an electrostatic potential drives the dynamics of individual electrons tunneling into a metallic island. We also discuss the role of coherence in gambling demons measuring quantum jump trajectories.
\end{abstract}

\maketitle

Maxwell's demon, as introduced in 1867~\cite{Maxwell}, is a little intelligent being who acquires information about the microscopic degrees of freedom of two gases held in  two containers at different temperatures, and separated by a rigid wall. The demon is able to control a tiny door, which can be opened {\em at stochastic times}, allowing fast particles from the cold container pass to the hotter one,  and hence generating a  heat current against a temperature gradient. This paradoxical behavior challenging the second law of thermodynamics, has its roots in the link between information and thermodynamics, which has fascinated scientists from more than a century~\cite{Review1}.
Maxwell's demon is nowadays considered a paradigmatic example of  feedback control, for which modified thermodynamic laws apply~\cite{MDlaws1, MDlaws2, Rio, MDlaws5} which have been tested experimentally in classical~\cite{MDexperimentsREV, bechhoefer,ritort} and quantum systems~\cite{MDQexperiments2, MDQexperiments3}.

\begin{figure}[t!]
 \includegraphics[width= \linewidth]{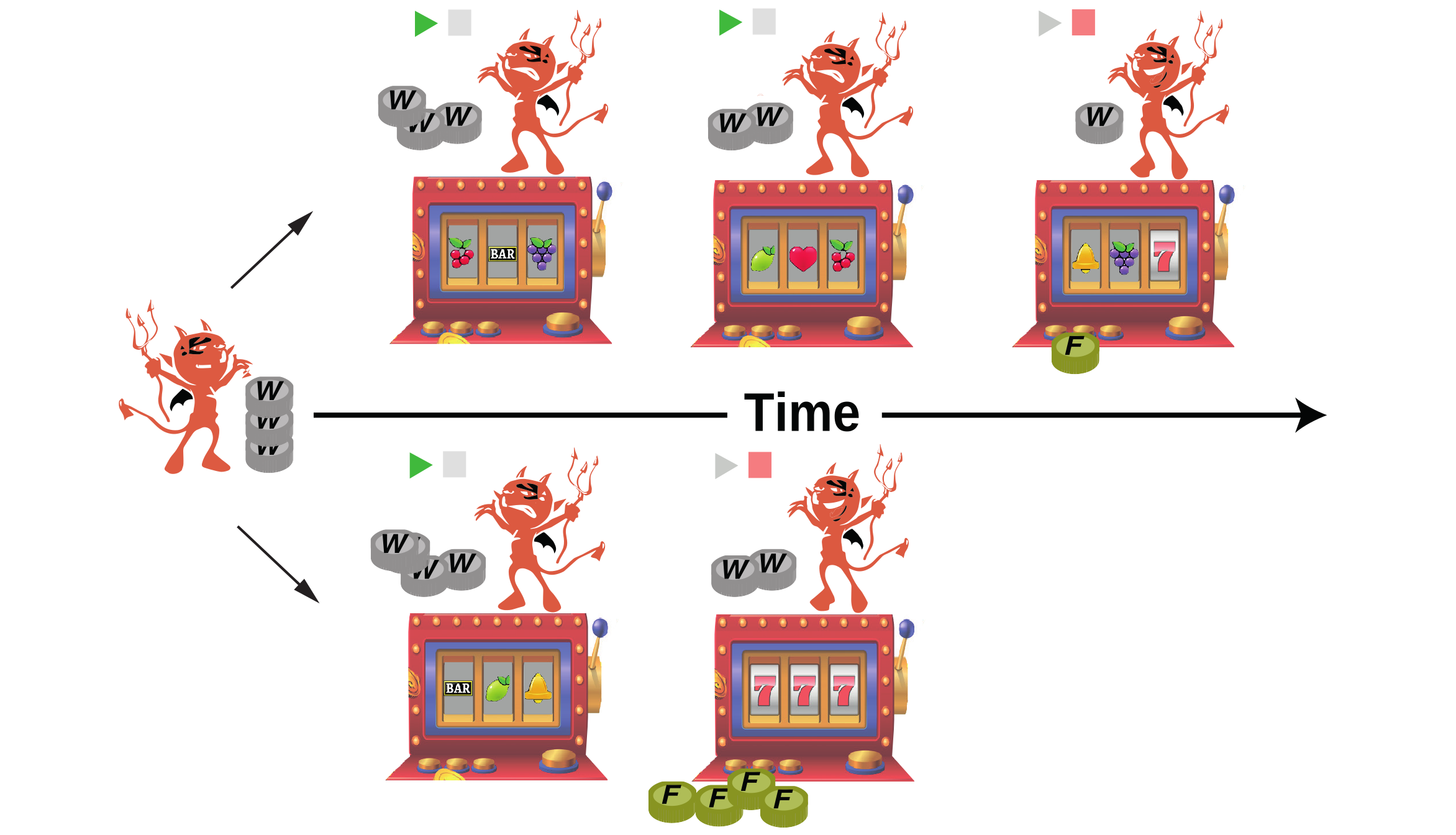}
 \caption{\textbf{Illustration of a gambling demon.} The demon spends work ($W$, silver coins) on a physical system (slot machine) hoping to collect free energy ($F$, gold coins) by executing a gambling strategy. In each time step, the demon does work on the system (introduces a coin in the machine) and decides whether to continue ("play" sign) or to quit  gambling and collect the prize ("stop" sign)  at a stochastic time $\mathcal{T}$  following a prescribed strategy. In the illustration, the demon plays the slot machine until a fixed time $\mathcal{T}=3$ (top row) unless the outcome of the game is beneficial at a previous time, e.g. $\mathcal{T}=2$ (bottom row). Under specific gambling schemes, the demon can extract on average more free energy than the work spent over many iterations, a scenario that is forbidden by the standard second-law inequality.\vspace{-0.5cm}} 
 \label{fig1}
\end{figure}

Here we propose and realize a {\em gambling demon} which can be seen as a variant of the original Maxwell's thought experiment (Fig.~\ref{fig1}). Such gambling demon invests work by performing a nonequilibrium thermodynamic process and acquires information about the response of the system during its evolution. Based on that information, the demon  decides whether to stop the process or not following a given set of stopping rules and, as a result,  may recover  more work from the system than what was invested. However, differently to Maxwell's demon, a gambling demon does not control the system's dynamics, hence excluding the possibility of  proper feedback control. This is analogous to a  gambler who invests coins in a slot machine hoping to obtain a positive payoff. Depending on the sequence of outputs from the slot machine, the gambler may decide to either continue playing or stop the game (e.g. to avoid major losses), according to some prescribed strategy. How much work may the gambling demon save/extract on average in a given transformation by implementing any possible strategy?

In this Letter, we  derive and test experimentally  universal equalities and inequalities for the  work and entropy production fluctuations in Markovian nonequilibrium processes subject to gambling strategies that stop the process at a finite time during an arbitrary deterministic driving protocol. Our  results apply to both classical and quantum stochastic dynamics, and provide tight bounds to work extraction beyond the generalized second laws with continuous feedback control~\cite{ritort}. We derive these results applying the theory of Martingale stochastic processes. Martingales have been fruitfully applied in probability theory~\cite{Williams}, quantitative finance~\cite{economy}, and more recently in nonequilibrium thermodynamics~\cite{raphaelshamik,neri,moslonka2020memory,ge2018martingale,yang2020unified}, providing  insights beyond standard fluctuation theorems, e.g. universal bounds for the  extrema and stopping-time statistics of thermodynamic quantities~\cite{neri, raphaelshamik, chetrite, ours, neri2, neriW}.

{\it \flushleft Work fluctuation theorems at stopping times---} We consider thermodynamic systems in contact with a thermal bath with inverse temperature $\beta= 1/ k_B T$. The Hamiltonian $H$ of the system depends on time through an external control parameter $\lambda(t)$ following a  prescribed deterministic protocol $\Lambda =\{ \lambda(t); 0\leq t \leq \tau \}$ of fixed duration $\tau$. The evolution of the system is subject to thermal fluctuations and thus we will describe its energetics using the framework of stochastic thermodynamics~\cite{sekimoto, SeifertREV,JarzynskiREV}. We denote the state (continuous or discrete) of the system  at time $0\leq t\leq \tau$ by  $x(t)$, and the probability of observing a given trajectory $x_{[0,\tau]} \equiv \{x(t)\}_{t=0}^{\tau}$ associated with the driving protocol $\Lambda$ by $P(x_{[0, \tau]})$.  We assume its dynamics is stochastic and Markovian with  probability density  $\varrho(x,t)$. Thermodynamic variables such as system's energy $E(t)=H(x(t),\lambda)$ and entropy $S(t) \equiv -k_{\rm B}\ln \varrho(x(t),t)$ are then stochastic processes,  functionals of the stochastic trajectories  $x_{[0,\tau]}$. We denote $W(\tau) \equiv \int_0^\tau dt~\partial_t H(x(t),t)$ the work exerted on the system up to time $\tau$, and $\Delta F(\tau) \equiv F(\tau) - F(0)$ the nonequilibrium free energy change, with $F(\tau) \equiv E(\tau) - T S(\tau)$.
A key result from stochastic thermodynamics is the  fluctuation theorem $\langle e^{-\beta (W-\Delta F)}\rangle=1$~\cite{jarz,seifert}, which implies the second-law inequality $\langle W \rangle - \langle \Delta F\rangle\geq 0$, where the averages $\langle\,\cdot\,\rangle$ are done over all possible trajectories of duration $\tau$  in the nonequilibrium protocol $\Lambda$.  

We now ask ourselves whether the work fluctuation theorem and the second law still hold when averaging over trajectories  stopped at stochastic times, following a custom ``gambling" strategy. 
We consider strategies defined through a generic \emph{stopping condition} that can be checked at any instant of time $t$ based only on the  information collected about the system up to that time. 
In each run, the demon gambles applying the prescribed stopping condition, and decides whether to stop gambling or not depending on the system's evolution. In this work, we consider stopping times obeying $\mathcal{T}(x_{[0,\tau]}) \leq \tau$ for any trajectory $x_{[0,\tau]}$, i.e. demons which are enforced to gamble before or at the end of the nonequilibrium driving.  For this class of systems we derive the inequality
\begin{equation}\label{eq:stopSL}
 \langle W \rangle_\T -  \langle \Delta F\rangle_{\T} \geq - k_B T\, \langle \delta \rangle_\mathcal{T},
\end{equation}
which involves  averages of functionals  of trajectories evaluated at stopping times $\langle O \rangle_\T = \sum_{x_{[0, \mathcal{T}]}} P(x_{[0, \mathcal{T}]}) O(\T)$, i.e. taken over many trajectories $x_{[0, \mathcal{T}]}$, each stopped at a stochastic time~$\mathcal{T}$.
Importantly, the quantity  
\begin{equation}\label{eq:ducal}
\delta(\mathcal{T}) \equiv \ln \left[ \frac{\varrho(x(\mathcal{T}),\mathcal{T})}{\tilde{\varrho}(x(\mathcal{T}),\tau-\mathcal{T})} \right],
\end{equation}
denoted here  as {\em stochastic distinguishability}, is  a trajectory-dependent  measure of how distinguishable is $\varrho(x, \mathcal{T})$ with respect to the probability distribution $\tilde{\varrho}( x, \tau - \mathcal{T})$ at the same stopping (i.e. stochastic) time $\mathcal{T}$ in a reference time-reversed process which is defined as follows.  Its driving protocol   $\tilde{\Lambda} = \{ \tilde{\lambda}(\tau - t); 0\leq t \leq \tau \}$  is the time-reversed picture of the forward protocol and its initial distribution is  the distribution obtained at the end of the forward protocol, i.e.  $\tilde{\varrho}(x, 0) \equiv \varrho(\tilde{x}, \tau)$~\cite{tildenote}. 
We derive Eq.~\eqref{eq:stopSL} by extending the Martingale theory of stochastic thermodynamics to generic driven Markovian processes  starting in arbitrary nonequilibrium conditions. This leads us to the  fluctuation relation at stopping times 
\begin{equation}\label{eq:stopJE}
\langle e^{-\beta (W- \Delta F)-\delta} \rangle_\mathcal{T} = 1,
\end{equation} 
which implies Eq.~\eqref{eq:stopSL} by Jensen's inequality~\cite{SM}. 
For the particular case of deterministic stopping at the end of the protocol $\T \rightarrow \tau$,  we get $\delta(\T) \rightarrow 0$ and thus Eqs.~\eqref{eq:stopSL} and~\eqref{eq:stopJE} recover respectively the standard second law and the work fluctuation theorem, as expected.

Equation~\eqref{eq:stopSL} reveals that the time-asymmetry introduced by the driving protocol, $\langle \delta \rangle_{\T} \geq 0$,  enables for an apparent ``second-law violation'' i.e.  $\langle W\rangle_\T\leq \langle\Delta F\rangle_\T$ at stopping times~\cite{apparent}. Because the system's evolution is stopped at stochastic times  at which the external protocol takes on different values, the average work done in the gambling process is not bounded by  the free energy change $ \langle \Delta F\rangle_\T$ between the initial and the final state  that one could reach with a deterministic protocol leading to the distribution  $\varrho(x,\T)$. The maximum extent of the violation of the traditional statement of the second law increases with $\langle\delta\rangle_{\T}$ i.e. when the process is driven far from equilibrium and the dynamics is strongly time asymmetric.
Eqs.~\eqref{eq:stopSL} and~\eqref{eq:stopJE} are valid for {\em any stopping strategy}, thereby  introducing a new level of universality. We next put to the test our results applying one specific set of stopping times to experimental data.

\begin{figure}
	\includegraphics[width=\linewidth]{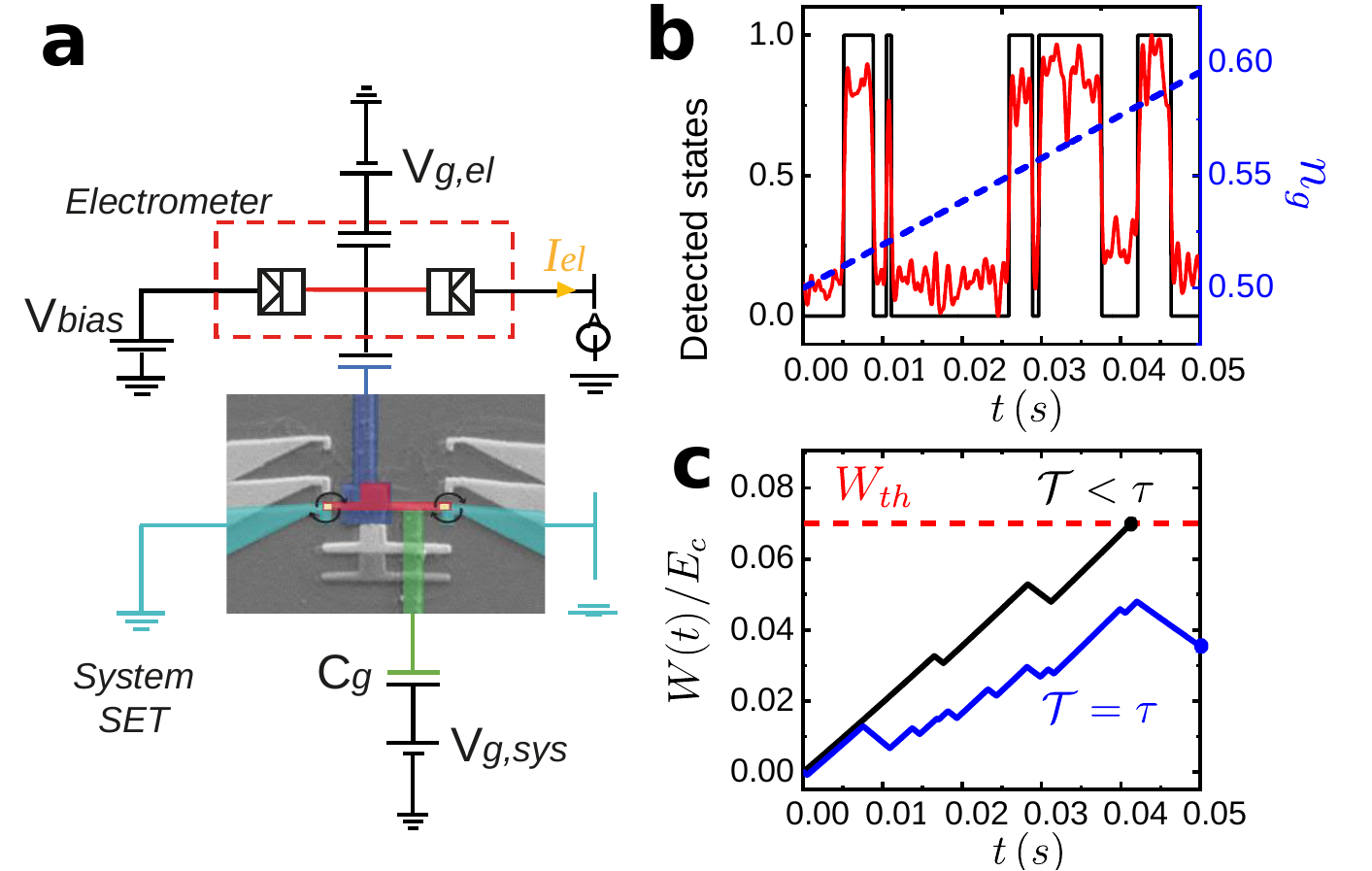}
	\caption{\textbf{a.}~Scanning electron micrograph of the  single-electron box (SEB) with false-color highlight on the Cu island (red) and the Al superconducting lead (turquoise). The superconducting leads are tunnel-coupled through thin oxide barriers (yellow) to the island. The DC SET electrometer is coupled capacitively to the box through a bottom electrode (blue) detects the excess charge of the box $n(t)$. \textbf{b.}~Representative time traces of the current measured through the electrometer (red solid line) and its digitized version (black solid line). The blue dashed line correspond to the driving protocol $n_g(t)$ of duration  $\tau = 0.05$s. \textbf{c.}~Example traces of the  stochastic work done on the box as a function of time. We execute the following gambling strategy: the process is stopped at $\mathcal{T}<\tau$ (black line) only when the work   reaches a threshold value $ W_{th}$  (red dashed line) before $\tau$. On the contrary, the process is stopped at final protocol time $\mathcal{T}= \tau$ if the work threshold is never reached during the driving protocol (blue line).} \label{fig:ExpSetup}
\end{figure}

{\it \flushleft  Experimental verification---} The experimental setup that we used to test the aforementioned predictions (shown in Fig.~\ref{fig:ExpSetup}\textbf{a}) consists of two capacitively-coupled metallic islands with small capacitance forming a single-electron transistor (SET) as a detector, and a single-electron box (SEB) as the system~\cite{expdata,SETs}.  The  SEB, with capacitance $C$, is left unbiased: the offset charge $n_g$ of the SEB can be externally tuned with a gate voltage $V_{g,sys}=en_g/C_g$.  At low temperature $k_BT<e^2/2C$ the box can be approximated as a two-state system with charge number states $n = 0$ and $n= 1$, and the offset charge tuning enables the control of individual electrons on the island through the change in its electrostatic energy $E_c(n-n_g)^2$, with $E_c = 1.94 k_B T$ and $T = 0.67 {\rm K}$. The other SET is used as an electrometer biased with a low voltage: through capacitive coupling to the box, its output current is sensitive to the box charge state, taking two values corresponding to the system states.
The  tunnelling of an electron into the island corresponds to a jump between the states $n = 0$ and $n= 1$ and is associated with an energy cost $\epsilon(n_g) = E_c(1-2n_g)$.  
 Through continuous monitoring of the box state $n(t)$ (see Fig.~\ref{fig:ExpSetup}\textbf{b}), we experimentally evaluate  at real time the heat exchange  between the system and the bath during a driving protocol of the gate voltage $n_g(t) = \lambda(t)$. The tunnelling (i.e. heat exchange)  events occur at rates of order~$\Gamma_d\sim 230$ Hz. If a jump occurs at time $t$ within a sampling time $\Delta t=20~\mu$s $\ll\Gamma_d^{-1}$ at gate voltage $n_g$, the work increment is $\delta W=0$ and the heat increment is $\delta Q = \epsilon(n_g)$  [$\delta Q = -\epsilon(n_g)$] for an electron tunneling into (out) of the island. Conversely, if no jump occurs, $\delta Q=0$ and $\delta W=2E_c(n_g-n)\dot{n}_g\delta t$.

\begin{figure*}[ht]
	\includegraphics[width=\linewidth]{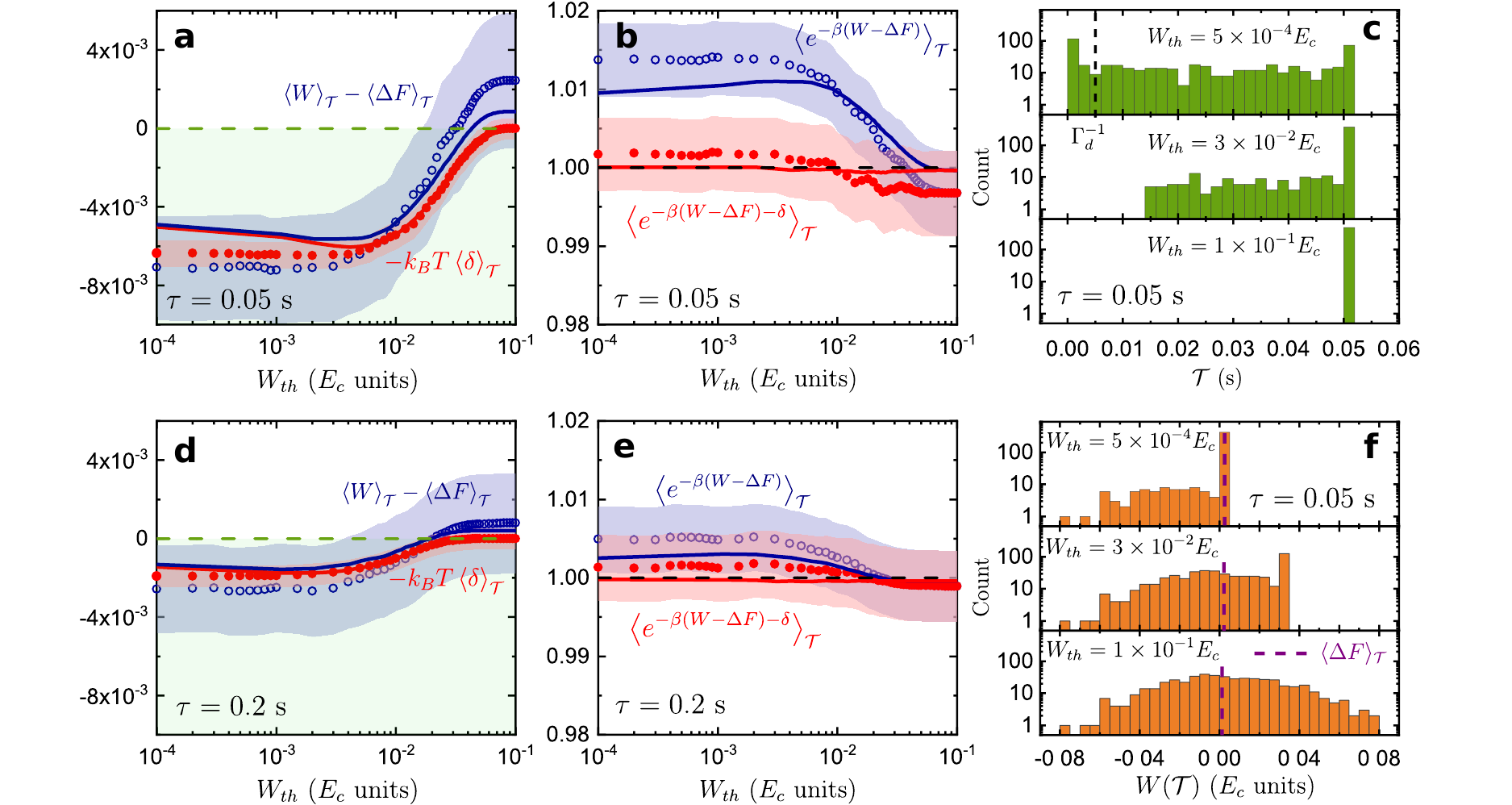}
	\caption{Dissipated work $\langle W\rangle_{\mathcal{T}} -\langle \Delta F\rangle_{\mathcal{T}}$ (blue) and stochastic indistinguishability at stopping times (red) $-k_B T\langle\delta(\mathcal{T})\rangle$ (dots: experimental data, solid lines: simulation) in charging energy $E_c=109{\rm \mu eV}$ units averaged over many realizations for protocol durations $\tau$= 0.05 s (a) and $\tau$ = 0.2 s (d) as a function of  work threshold values. \textbf{b,e}. test of the generalized work fluctuation relation and of Eq.~\eqref{eq:stopJE}  (dots: experimental data, solid lines: simulation) for $\tau$ = 0.05 s (\textbf{b}) and  $\tau$ = 0.2 s ({\bf e}). {\bf c,f}. Distributions of stopping times $\mathcal{T}$ ({\bf c}) and corresponding work values $W(\mathcal{T})$ ({\bf f}) for a ramp time $\tau=$ 0.05 s for work thresholds $W_{th}=5\times 10^{-4},3\times 10^{-2}$ and $10^{-1}E_c$. The total uncertainty is shown by shadowed areas; it is the combination of the statistical uncertainty and error on temperature (about 10\%).}\label{fig:result}
\end{figure*}

The experimental driving protocol $ \Lambda$ of duration $\tau$ is depicted in Fig. \ref{fig:ExpSetup}\textbf{b}. The system is initially prepared at charge degeneracy, i.e., $n_g(0)=1/2$ at thermal equilibrium where the initial energies of states  are equal, following a uniform distribution. Then the energy splitting is tuned according to a linear ramp, $\epsilon  [n_g(t)] = 1/2+\Delta n_gt/\tau$, with $\Delta n_g=0.1$ fixed throughout the experiment.  The protocol is repeated several times ($\sim 500-1000$) to acquire sufficient statistics. 
The gambling strategy that we chose consists on stopping the dynamics at stochastic times $\mathcal{T}$  when the work exceeds a threshold value $W_{th}$ (red dashed line) or at $\tau$ otherwise. 
The gambling strategy was applied \textit{a posteriori} on the data: for the same set of traces taken for the full protocol duration, the stopping condition (threshold work $W_{th}$) was varied between $10^{-4}E_c$ and $10^{-1}E_c$.
In Fig. \ref{fig:ExpSetup}\textbf{c} we present two examples of stopped work trajectories  where one reaches the threshold value at a time $\mathcal{T} < \tau$ (black line), while the other remains below the threshold until the final time $\tau$ (blue line).

Experimental values of $\langle W \rangle_\T -\langle \Delta F\rangle_\T$ and $-k_B T\langle\delta\rangle_\T$ are shown in Figure \ref{fig:result}\textbf{a} and \ref{fig:result}\textbf{d} for two different ramps of durations  $\tau = 0.05$s (\textbf{a}) and $\tau= 0.2$s (\textbf{d}) as a function of the  work threshold $W_{th}$.  These results are validated and are in good agreement with  numerical simulations over the entire  threshold range when including the experimental uncertainty.
For both   ramp durations $\langle W \rangle_\T -\langle \Delta F \rangle_\T$ is negative at small $W_{th}$,
defying the conventional second law but is yet in agreement with  Eq.~\eqref{eq:stopSL} within experimental errors. We find that the faster is the protocol, the more negative  $\langle W \rangle_\T -\langle \Delta F \rangle_\T$ becomes, which can be understood as a consequence of the irreversibility (and hence $\langle \delta \rangle_\T$) associated with the  ramp driving  speed. 
For large values of  $W_{th}$,  almost all trajectories are stopped at $\tau$ and the conventional second law is recovered, as $\langle\delta\rangle_{\mathcal{T}}$ becomes small. Furthermore, Figs.~\ref{fig:result}\textbf{b} and \textbf{e} report the exponential averages  $\langle e^{-\beta (W- \Delta F)} \rangle_\mathcal{T}$ and $\langle e^{-\beta (W- \Delta F)-\delta} \rangle_\mathcal{T}$ evaluated at the stopping times.  Notably, the conventional work fluctuation theorem $\langle e^{-\beta(W - \Delta F)}\rangle_{\mathcal{T}}=1$ only holds for large $W_{th}$, while for small  $W_{th}$, $\langle e^{-\beta(W - \Delta F)}\rangle_{\mathcal{T}}$ is  significantly greater than one within experimental errors. On the other hand, we obtain an excellent agreement (with accuracy $\sim99.5$\%) of our fluctuation relation~\eqref{eq:stopJE} for all values of $W_{th}$ and both ramp speeds. To gain further insights, in Figs.~\ref{fig:result}\textbf{c} and~\ref{fig:result}\textbf{f} we show histograms of the stopping times $\mathcal{T}$ and the value of the work at the stopping time $W(\mathcal{T})$.  For small thresholds we observe that the  distribution of $\mathcal{T}$ is broad and includes stopping events that take place at short times $\mathcal{T} \lesssim \Gamma_d^{-1}$ (Fig.~\ref{fig:result}\textbf{c}, top panel). Its corresponding  distribution of $W(\mathcal{T})$ (Fig.~\ref{fig:result}\textbf{f}, top panel) has a peak at $W_{th}$  arising from trajectories stopped before $\tau$ and a  tail  $W(\mathcal{T})<\langle\Delta F\rangle_{\mathcal{T}}$  from trajectories ending at the end of the protocol. By increasing the threshold value (Fig.~\ref{fig:result}\textbf{c} and~\ref{fig:result}\textbf{f},  middle panels) we reduce the number of trajectories that stop before $\tau$ hence the distribution of $\mathcal{T}$ becomes narrower (Fig.~\ref{fig:result}\textbf{c}, bottom panel). This effect is accompanied by a broadening of the  $W(\mathcal{T})$  distribution recovering a Gaussian-like shape with mean above the free energy change for  large enough $W_{th}$ (i.e. typically far outside the standard fluctuation interval of $W$), Fig.~\ref{fig:result}\textbf{f}  bottom panel.

{\it \flushleft Quantum gambling---} The gambling demon can also be extended to the quantum realm by considering quantum jump trajectories~\cite{milburn}. Here the pure state of the system $\ket{\psi (t)}$ follows stochastic evolution conditioned on the measurement outcomes generated by the continuous monitoring of the environment~\cite{hekking, horowitz, manzano}. 

In this case,  we derive the following quantum stopping-time work fluctuation relation 
\begin{equation}\label{eq:stopFTq}
\langle e^{-\beta[W - \Delta F] - \delta_\mathrm{q} + \Delta S_\mathrm{unc}} \rangle_\T = 1,
\end{equation}
where again $W$ and $\Delta F$ are respectively the work performed and free energy change during trajectories stopped at $\mathcal{T}$~\cite{SM}. The term $\delta_\mathrm{q}(t) \equiv \ln \langle \psi(t) | \rho(t) |\psi(t) \rangle - \ln \langle \psi(t) | \Theta^\dagger \tilde{\rho}(\tau - t) \Theta |\psi(t) \rangle$ is the quantum analogue of Eq.~\eqref{eq:ducal}, $\rho$ and $\tilde{\rho}$ being the density operators in the forward and backward process respectively, and $\Theta$ the time-reversal operator in quantum mechanics. 
As before, time-inversion at time $\tau$ implies $\delta_\mathrm{q}(\tau) = 0$. The key difference of the quantum fluctuation relation \eqref{eq:stopFTq} with respect to its classical counterpart in Eq.~\eqref{eq:stopJE} is the appearance of a genuine entropic term associated to quantum measurements, namely the ``uncertainty''  entropy production
\begin{equation}\label{eq:sunc}
\Delta S_\mathrm{unc}(\mathcal{T}) = - \ln \left( \frac{ \langle n(\mathcal{T})| \rho(\mathcal{T}) | n(\mathcal{T}\rangle)  }{\langle \psi(\mathcal{T})| \rho(\mathcal{T}) | \psi(\mathcal{T}) \rangle} \right).
\end{equation}
This quantity measures how much more surprising is a particular eigenstate $|n(t)\rangle$ of $\rho(t)$ with respect to the stochastic wave function $\ket{\psi(t)}$, as characterized by the logarithm of the squared Uhlman fidelity, $\langle \psi(t) | \rho(t) |\psi(t) \rangle$~\cite{ours}.  In general, $\ket{\psi(t)}$ can be an arbitrary superposition of the instantaneous eigenstates $\ket{n(t)}$.  In the classical limit  the stochastic evolution of $\ket{\psi(t)}$ is given by jumps between energy levels and thus $\ket{\psi(\mathcal{T})} = \ket{n(\mathcal{T})}$. Consequently $\Delta S_\mathrm{unc}(\mathcal{T}) = 0$ in Eq.~\eqref{eq:sunc} and $\delta_\mathrm{q}(\mathcal{T}) = \delta(\mathcal{T})$ for any $\mathcal{T}$, thus recovering Eq.~\eqref{eq:stopJE} in the classical limit.
The corresponding stopping-time second-law inequality for quantum systems reads
$\langle W \rangle_\T -  \langle \Delta F \rangle_\T \geq  - k_B T (\langle \delta_\mathrm{q} \rangle_\T - \langle \Delta S_\mathrm{unc} \rangle_\T )$,
where $\langle \Delta S_\mathrm{unc} \rangle_\T$ modifies the entropic balance. Even if $\langle \Delta S_\mathrm{unc} \rangle \geq 0$ for any fixed time $t\leq \tau$, the average over stopped trajectories $\langle \Delta S_\mathrm{unc} \rangle_\T$ may be either positive or negative depending on the selected gambling strategy. Therefore, the quantum fluctuations induced by measurements 
may act either as an entropy source or as an entropy sink.

{\it \flushleft Conclusions---} We have introduced and illustrated the stochastic thermodynamics  of  gambling demons, i.e. driven nonequilibrium processes that are stopped at stochastic times following a prescribed criterion.
Our results generalize the second law to arbitrary stopping (``gambling") strategies for classical and quantum systems driven out of equilibrium. 
Even though all finite-time horizon gambling strategies fulfill the stopping-time fluctuation relation~\eqref{eq:stopJE} and the inequality~\eqref{eq:stopSL}, not all guarantee average work extraction above  the average nonequilibrium free energy change. Such ``negative dissipation" requires the usage of gambling strategies in a sufficiently irreversible process:  stopping the dynamics at stochastic times with a suitable gambling strategy, and a time-asymmetric driving protocol. This contrasts with heat and information engines which achieve maximal work extraction in the  quasistatic reversible limit~\cite{martinez2016brownian,horowitz2011thermodynamic}. 

Our relations are fundamentally different to the generalized second law with feedback $\langle W\rangle - \langle \Delta F \rangle \geq - k_B T\,I $, where $I$ is the information acquired by a feedback controller from the system in a fixed-time protocol~\cite{MDlaws1, MDlaws2}, or at stochastic times~\cite{ritort}.
The information used to implement a gambling strategy can be estimated assuming periodic measurements every sampling time $\Delta t$, each providing at least a bit of information, correspoding to ``stop''/``don't stop'' the trajectory. In the small sampling time limit, these measurements generate sequences of $N \sim \mathcal{T}/\Delta t + 1$ bits per trajectory. Erasing these bits would have an energetic cost that becomes infinitely large in the continuous measurement limit $\Delta t \rightarrow 0$~\cite{ritort}.  
 Our results show that gambling demons are, nevertheless, constrained by the bound in Eq.~\eqref{eq:stopSL}, which is tighter than an extension of the second law with feedback at stopping times. 
 In the experiment reported here we indeed obtain $k_B T \langle \delta \rangle_\mathcal{T} \sim 7.8 \times 10^{-3} k_B T \ll kT \ln2$, but faster protocols are expected to achieve larger values of $\langle \delta \rangle_\mathcal{T}$. It would be interesting in the future to further investigate the interplay between our fluctuation relations and information acquisition, as well as with recent stopping-time uncertainty relations~\cite{PhysRevLett.125.120604}, and speed limits~\cite{shiraishi2018speed}.
Applications to experimental quantum devices~\cite{minev2019catch, Murch} may allow to exploit quantum superpositions to enhance work extraction beyond the classical limits. Finally, it would be interesting to explore optimization of stopping strategies using knowledge in quantitative finance (e.g. option pricing, arbitrage, etc.) and gambling~\cite{Dinis,Ito} such as Parrondo games~\cite{Harmer}.

\begin{acknowledgments}
We acknowledge fruitful discussions with Christopher Jarzynski. G.M. acknowledges funding from the European Union's Horizon 2020 research and innovation programme under the Marie Sk\l{}odowska-Curie grant agreement No. 801110 and the Austrian Federal Ministry of Education, Science and Research (BMBWF). R.F. research has been conducted within the framework of the Trieste Institute for Theoretical Quantum Technologies (TQT). This work was funded through Academy of Finland Grant No. 312057 and from the European Union’s Horizon 2020 research and innovation programme under the European Research Council (ERC) programme. 
\end{acknowledgments}

\clearpage

\onecolumngrid

\widetext
\begin{center}
\textbf{\large Supplemental Material to ``Thermodynamics of Gambling  Demons''}
\end{center}

\setcounter{equation}{0}
\setcounter{figure}{0}
\setcounter{table}{0}
\setcounter{page}{1}
\makeatletter
\renewcommand{\thesection}{S\arabic{section}}
\renewcommand{\theequation}{S\arabic{equation}}
\renewcommand{\thefigure}{S\arabic{figure}}
\renewcommand{\bibnumfmt}[1]{[S#1]}
\renewcommand{\citenumfont}[1]{S#1}

\newcommand{\Li}{\mathcal{L}}
\def \id {{\mathds 1}}

\

This Supplemental Material consist in two parts. The first part, corresponding to Sec.~\ref{sec:DM}, is dedicated to introduce further details on the methods, both experimental and theoretical ones, used in the main text. In the second part, Sec.~\ref{sec:proofs}, we provide rigorous proofs for the main theoretical results, including Eqs.~(1),(3) and (4) of the main text.

\section{Detailed Methods} \label{sec:DM}
In this section we provide details on the methods used to obtain the results reported in the main text. In particular, in Sec.~\ref{sub:exp} we provide further experimental details regarding the characterization of the single-electron box and its occupation probabilities. In Sec.~\ref{sub:qtraj} we review the main elements of quantum jump trajectories used in this work, while Sec.~\ref{sub:termo} is devoted to introduce the quantum thermodynamic framework and quantities of interest. An extended discussion on the non-trivial consequences that the presence of coherence introduce when stopping strategies are applied is provided in Sec.~\ref{sub:stop}. In Sec.~\ref{sub:martinq} we give details on the quantum martingale theory used to obtain the main results in the text, together with its classical limit that we explicitly obtain in Sec.~\ref{sub:martinc}.     

\subsection{Experimental details} \label{sub:exp}
The single-electron box can be conveniently approximated as a classical two-level system (with states "0" or "1" extra electron on the island), with internal energy $U=E_c(n-n_g)^2$ a gate-tunable energy splitting $\epsilon(n_g)=U(1,n_g)-U(0,n_g)=E_c(1-2n_g)$ corresponding to the energy cost for an electron to tunnel into the box. The charging energy $E_c=e^2/2C=109~\mu$eV, with $C$ the box capacitance, is extracted by direct current-voltage measurements (Coulomb diamonds, see \cite{Sexpdata} and supplementary material within). The system's temperature is extracted by simply measuring a time trace containing a statistically significant number of tunneling events at fixed $n_g$ (i.e. at equilibrium) for $0\leq n_g\leq 1$. Since the state-space of the system is discrete, here we use for convenience the notation $p_{n(t)}(t) \equiv \varrho(n(t),t)$. The occupation probabilities at equilibrium follow the detailed balance $p_0^{\mathrm{eq}}/p_1^{\mathrm{eq}}=e^{\epsilon(n_g)/k_BT}$, a property that we use to extract the effective temperature $T=670$ mK \cite{Sexpdata}. The tunneling rates can be generally derived using Fermi's Golden Rule and the so-called orthodox theory of electron tunneling \cite{PhysRevB.44.6199}. Close to charge degeneracy ($n_g=1/2$), for which $\epsilon=0$, these rates can be rather well approximated by the following expression:
\begin{equation}
\label{rates_simple}
\Gamma_{1\rightarrow 0\,(0\rightarrow 1)}(n_g)\approx\Gamma_d\exp\left[\mp\frac{\epsilon(n_g)}{2k_BT}\right],
\end{equation}
where $\Gamma_d\approx 230$ Hz is the experimentally determined \cite{Sexpdata} tunneling rate at charge degeneracy. When driven out of equilibrium by a short linear ramp $\Lambda(t) = \{n_g(t);0\leq t \leq \tau \}$ with $n_g(t)=1/2+\Delta n_g/\tau$, the occupation probabilities obey a standard, protocol-dependent master equation:
\begin{equation}
  \left\{
    \begin{aligned}
      \dot{p}_1(t) &=-\Gamma_{1\rightarrow 0}[n_g(t)]p_1(t) +\Gamma_{0\rightarrow 1}[n_g(t)]p_0(t)\\
      \dot{p}_0(t) &=-\Gamma_{0\rightarrow 1}[n_g(t)]p_0(t) +\Gamma_{1\rightarrow 0}[n_g(t)]p_1(t)\\
    \end{aligned}
  \right.
  \label{OOE_mastereq}
\end{equation}
We numerically solve this system with initial conditions $p_0(0)=p_1(0)=1/2$, and the parameters used in the experiment ($\Delta n_g=0.1$, a discrete time step $\Delta t=20\,\mu$s corresponding to the data digitization rate, and the protocol times $\tau=50$ ms, $200$ ms). The experimental out-of-equilibrium probabilities are reconstructed from the traces: for each time instant we simply take the average of the measured state over the repetitions.
\begin{figure}
	\centering
	\includegraphics[width=8.7cm]{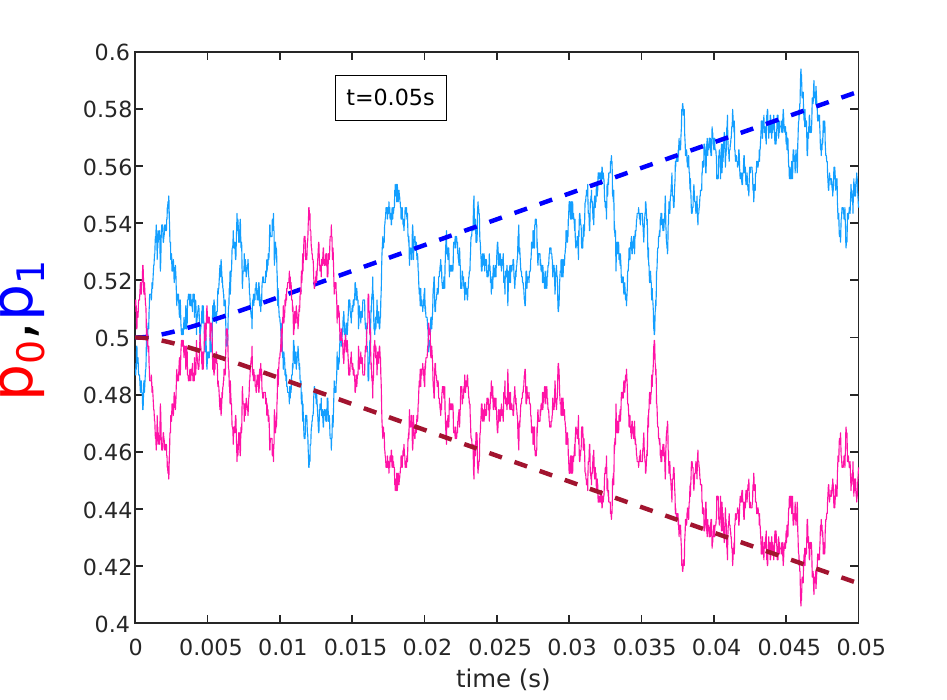}
	\includegraphics[width=8.7cm]{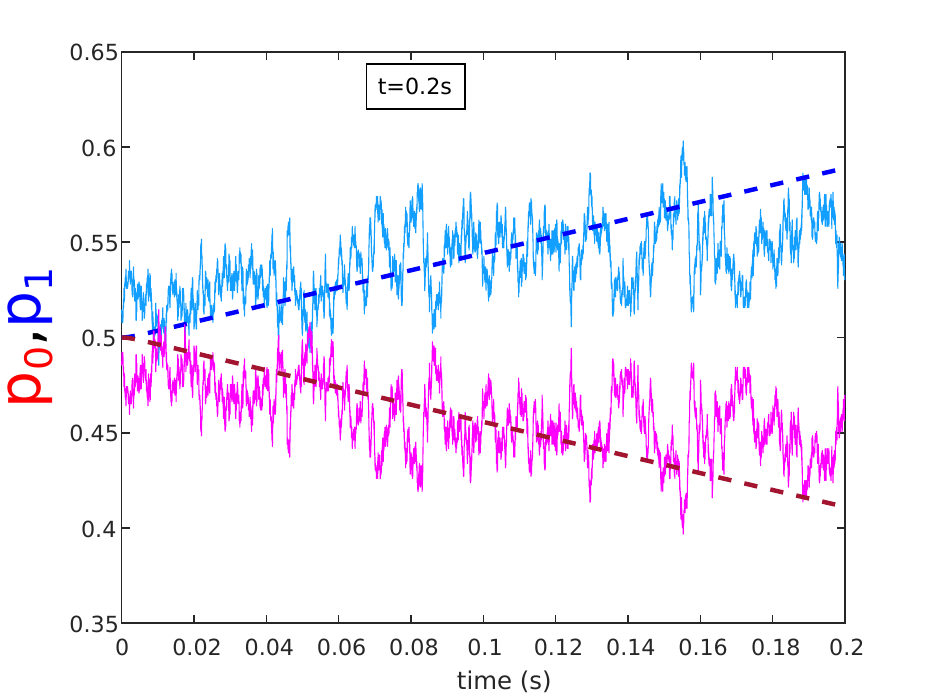}
	\caption{Out-of-equilibrium occupation probabilities of the box states under the ramp driving $\Lambda(t)$ from experimental traces ($\sim 1000$). The dashed lines are the numerical solutions of the probabilities using Eq.~(\ref{OOE_mastereq}).}
	\label{OOE_proba}
\end{figure}

The experimental occupation probabilities are shown in Fig. \ref{OOE_proba}. The resolution is limited by the relatively low number of experimental traces ($\sim 1000$), but the data are in fair agreement with the solutions from Eq. (\ref{OOE_mastereq}). Based on this agreement, we use the numerical solutions to evaluate the stochastic free energy difference at a given experimental stopping time: $\Delta F(\mathcal{T})=\Delta U(\mathcal{T})+k_BT\ln[p_{n(\mathcal{T})}(\mathcal{T})/p_{n(0)}(0)]$, where $\Delta U(\mathcal{T})=U(\mathcal{T})-U(0)$. In addition, we use the master equation to obtain the probabilities $\tilde{p}_{0,1}(t)$ associated with backward trajectories under the time reversed protocol $\tilde{\Lambda}=\{n_g(\tau-t); 0\leq t \leq \tau\}$, using the initial condition $\tilde{p}_{0,1}(0)=p_{0,1}(\tau)$. The obtained solutions allow us to evaluate the stochastic distinguishability term at stopping times $\delta(\mathcal{T})=\ln[p_{n(\mathcal{T})}(\mathcal{T})/\tilde{p}_{n(\mathcal{\mathcal{T}})}(\tau-\mathcal{T})]$. 



\subsection{Quantum jump trajectories} \label{sub:qtraj}
In order to describe Markovian stochastic quantum dynamics, we use the formalism of quantum jump trajectories~\cite{Smilburn}. This framework allows to describe the evolution of a pure state of the system, $|\psi(t) \rangle$, conditioned on a set of outcomes retrieved from continuous monitoring of the environment. The evolution consist in periods of smooth dynamics intersected by quantum jumps occurring at random times, which produce abrupt changes in the state of the system. The occurrence of such jumps is linked to the exchange of excitations between system and reservoir (e.g. emission and absorption of photons) captured by the detector. Such dynamics is described by the Stochastic Schr\"odinger equation $(\hbar = 1)$:
\begin{eqnarray} \label{eq:stochastic}
 \mathrm{d} | \psi (t) \rangle &=\mathrm{d}t \left( -i H(\lambda)  + \sum_k\frac{\langle L_k^\dagger  L_k  \rangle_{\psi (t)} - L_k^\dagger  L_k}{2}\right)\! |\psi (t) \rangle \nonumber \\
 &+ \sum_k \mathrm{d}N_k(t) \left( \frac{L_k }{\sqrt{\langle L_k^\dagger L_k \rangle_{\psi(t)}}} - {\mathds 1} \right)\!|\psi(t)\rangle.
\end{eqnarray}
Here $H(\lambda)$ is a Hermitian operator (usually the system Hamiltonian), and the operators $L_k(\lambda)$ for $k = 1 ... K$ are the Lindblad (or jump) operators, both of which may depend on the control parameter $\lambda(t)$ following the driving protocol $\Lambda = \{\lambda(t) ; 0 \leq \lambda \leq \tau \}$ up to time $\tau$. The random variables $\mathrm{d}N_k(t)$ are Poisson increments associated to the number of jumps $N_k(t)$ of type $k$ detected up to time $t$ in the process. This variables take most of the time the value $0$, and they become $1$ only at specific times $t_j$ when a jump of type $k_j$ is detected in the environment. Here we denoted $\langle A \rangle_{\psi (t)} \equiv \langle \psi(t)| \,A \,|\psi(t) \rangle$ the quantum-mechanical expectation values, and ${\mathds 1}$ the identity matrix.

Recording the different type of jumps occurring during the stochastic dynamics and the times at which they were detected, one may construct a measurement record $\mathcal{R}_0^\tau = \{(k_1, t_1),...,(k_J,t_J) \}$, where $(k_j , t_j)$ denotes a jump of type $k_j$ observed at time $t_j$, where $j= 1, ... , J$ for a total number of jumps $J$, and $0 \leq t_1 \leq t_2 \leq ... \leq t_J \leq \tau$. If the average over many processes is taken, the evolution reduces to a Markovian process for the density operator of the system $\rho(t)$, ruled by a Lindblad master equation~\cite{SLindblad}
\begin{equation} \label{eq:master}
 \dot{\rho}(t) = -i [H, \rho(t)] + \sum_k L_k \rho(t) L_k^\dagger - \frac{1}{2} \{L_k^\dagger L_k, \rho(t) \}.
\end{equation}
In the case of a thermal environment all jumps occur in the energy basis, leading to the exchange of discrete energy packets $E_k(\lambda)$ with the environment that can be interpreted as heat~\cite{Shorowitz, Shekking}. When the Hamiltonian has a fixed basis during all the control protocol $\Lambda$, a classical Markovian process is recovered. In this case, taking only the diagonal elements of $\rho$ in the energy basis, we recover from Eq.~\eqref{eq:master} a classical master equation.

\subsection{Quantum stochastic thermodynamics} \label{sub:termo} 
The framework of quantum jump trajectories is particularly well suited for extending stochastic thermodynamics to the quantum realm~\cite{Shorowitz, Shekking, Shorowitz2, Sleggio, Scampisi, SmanzanoPRE, Sgong, Sliu, Selouard, Smanzano, Skarimi}.
An important feature of quantum setups is the need to place the driven processes within a two-measurements scheme. Here the system is subjected to projective measurements in the density operator eigenbasis both at the beginning [$\rho(0)$] and at the end [$\rho(\tau)$] of the protocol $\Lambda$. Therefore, in a trajectory $\gamma_{\{ 0, \tau\}} \equiv \{(n(0),0); \mathcal{R}_0^\tau; (n(\tau),\tau)\}$ the system is prepared in an eigenstate $| n(0) \rangle$ with probability $p_{n(0)}(0)$ in the first measurement. Then the state $|\psi(t)\rangle$ evolves from $t=0$ up to time $t=\tau$ according to a given environmental measurement record $\mathcal{R}_0^\tau = \{(k_1, t_1),...,(k_J,t_J) \}$, where jump processes $k_j$ were detected at stochastic times $t_j$. Finally, the system is projected in $|n(\tau)\rangle$ in the second measurement.
The changes in observables of the system such as energy and stochastic entropy are given by $\Delta E(\tau) = \langle n(\tau) |H(\lambda(\tau))| n(\tau) \rangle - \langle n(0) |H(\lambda(0))| n(0) \rangle$, and $\Delta S(\tau) = -\ln p_{n(\tau)}(\tau) + \ln p_{n(0)}(0)$, with $p_{n(\tau)}(\tau)= \langle n(\tau)|\rho(\tau) |n(\tau)\rangle$ and $p_{n(0)}(0)=\langle n(0)| \rho(0) |n(0)\rangle$ the eigenvalues of $\rho(\tau)$ and $\rho(0)$, respectively. Averaging these quantities  over many trajectories we recover the standard expressions for the  energy change $\langle \Delta E(\tau) \rangle = \mathrm{Tr}[H(\lambda(\tau) \rho(\tau)] - \mathrm{Tr}[H(0) \rho(0)]$ and von Neumann entropy change of the system $\langle \Delta S(t) \rangle = - \mathrm{Tr}[ \rho(\tau)\ln \rho(\tau)] +  \mathrm{Tr}[ \rho(0) \ln \rho(0)]$.

A key quantity measuring the irreversibility of the physical process along single trajectories is the stochastic entropy production
\begin{equation}\label{eq:ep}
 \Delta S_\mathrm{tot}(\tau) = \ln \frac{P (\gamma_{\{0,\tau\}})}{ \tilde{P}(\tilde{\gamma}_{\{0,\tau\}})} =  \Delta S(\tau) +  \sum_{j = 1}^J \Delta S_{\rm env}^{k_j},
\end{equation}
where $P(\gamma_{\{ 0, \tau\}})$ is the probability that trajectory $\gamma_{\{0,\tau\}}$ is generated, and $\tilde{P}(\tilde{\gamma}_{\{0,\tau\}})$ is the probability to obtain the time-reversed trajectory $\tilde{\gamma}_{\{0,\tau\}} = \{ n(\tau); \tilde{R}_\tau^0; n (0) \}$ in the time-reverse or backward process. In the backward process, the time-reversed protocol $\tilde{\Lambda}$ is implemented over the (inverted) final state of the system in the forward process, $\tilde{\rho} \equiv \Theta \rho(\tau) \Theta^\dagger$. The term $\Delta S_{\rm env}^{k_j}$ in Eq.~\eqref{eq:ep} is the environmental entropy change due to the jump $k_j$~\cite{Smanzano}. 
The stochastic entropy production obeys the integral fluctuation theorem $\langle e^{-\Delta S_\mathrm{tot}(\tau)} \rangle = 1$, leading to the second law inequality $\langle \Delta S_\mathrm{tot} (\tau) \rangle \geq 0$, where here the average is taken over complete trajectories $\gamma_{\{0, \tau\}}$. 

In the case of a driven system in contact with a single thermal reservoir at temperature $T$, we have $\sum_j \Delta S_{\rm env}^{k_j} = -Q(\tau)/T$, where $Q(\tau)$ is the heat realeased by the reservoir during the trajectory. In such case the entropy production reads:
\begin{equation} \label{eq:epw}
\Delta S_\mathrm{tot}(\tau) = \beta \left[W(\tau) - \Delta F(\tau) \right],
\end{equation}
where $W(\tau) = \Delta E(\tau) - Q(\tau)$ is the stochastic work performed during the trajectory, and $\Delta F(\tau) = \Delta E(\tau) - k_B T \Delta S(\tau)$ the non-equilibrium free energy change.

\subsection{Stopping quantum trajectories} \label{sub:stop}
The introduction of the two-measurements scheme has non-trivial consequences for the thermodynamic behavior of the system when gambling strategies are to be employed to stop the process. The reason is that thermodynamic quantities like work or free energy are only well defined once the second measurement in the scheme has been performed, which requires performing the second measurement at the time at which the trajectory is stopped. However, if the trajectory is stopped before the end of the protocol, the introduction of a projective measurement at any time $t \leq \tau$ may disturb the trajectory. A quantum gambling demon willing to decide to stop or not the process at $\mathcal{T}$ must take the decision before the second measurement is performed, since otherwise quantum Zeno effect will trivialize the whole evolution. 
Therefore, the gambling demon decides to stop or not at $\mathcal{T}$ according to a selected stopping condition based on the information $\{(n(0),0); \mathcal{R}_0^\mathcal{T}\}$. If he stops, then the final measurement is performed in the $\rho(\mathcal{T})$ eigenbasis, completing the stopped trajectory $\gamma_{\{0,\mathcal{T}\}} = \{(n(0),0); \mathcal{R}_0^\mathcal{T}, (n(\mathcal{T}), \mathcal{T})\}$, otherwise the measurement is not performed and the evolution continues. This process introduces a final unavoidable disturbance of quantum nature in the stopped trajectories, that the gambling demon is not able to predict and/or control, with thermodynamic consequences.

In order to handle the thermodynamics of the measurement disturbance, we use the following  decomposition of the stochastic entropy production in Eq.~\eqref{eq:ep}:
\begin{equation} \label{eq:split}
\Delta S_\mathrm{tot}(t) = \Delta S_\mathrm{unc}(t) + \Delta S_\mathrm{mar}(t).
\end{equation}
Here the first term is the ``uncertainty'' entropy production already introduced in Eq.~(5) of the main text, and we denote the second term in Eq.~\eqref{eq:split} as the ``martingale'' entropy production:
\begin{equation} \label{eq:smar}
\Delta S_\mathrm{mar}(t) = - \ln \left( \frac{\langle \psi(t) | \rho(t)|\psi(t) \rangle}{\langle n(0) | \rho(0)|n(0) \rangle} \right)  + \sum_{j = 1}^J \Delta S_{\rm env}^{k_j}.
\end{equation}
This quantity represents a ``classicalization'' of the stochastic entropy production~\eqref{eq:ep}, containing a slightly modified boundary term which gets ride of the final projective measurement impact (first term), and the full extensive part due to the environmental entropy fluxes (second term). 

\subsection{Quantum Martingale theory} \label{sub:martinq} 
Our results for classical work fluctuation relations at stopping times derive from a more general martingale theory for entropy production that applies to both quantum and classical thermodynamic systems. This theory relates irreversibility, as measured by entropy production, in generic nonequilibrium processes with the remarkable properties of martingales processes. 

A martingale process is a stochastic process defined on a probability space whose expected value at any time $t$ equals its value at some previous time $s<t$ when conditioned on observations up to that time $s$. More formally, $M(t)$ is a martingale if it is bounded $\langle M(t) \rangle < \infty$ for all $t$, and verifies $\langle M(t) | M_{\{0,s\}} \rangle = M(s)$, where the later average is conditioned on all the previous values  $M_{\{0,s\}}$ of the process up to time $s$~\cite{SWilliams}.

We consider conditional averages of entropy production over trajectories with common history up to a certain time $t \leq \tau$ before the end of the protocol $\Lambda$, which constitutes the key ingredient for developing a martingale theory~\cite{Sneri,Sraphaelshamik}. We introduce the conditional average of a generic stochastic process $O(t)$ defined along a trajectory $\gamma_{\{0,t\}}$ as $\langle O(\tau) | \gamma_{[0,t]} \rangle = \sum_{n(\tau) , \mathcal{R}_t^\tau} O(\tau) P(\gamma_{\{0, \tau\}} | \gamma_{[0,t]})$, where the condition is made with respect to the {\it ensemble} of trajectories $\gamma_{[0, t]} \equiv \bigcup_{s=0}^t \gamma_{\{0,s\}}$ including all outcomes of  trajectories eventually stopped at all intermediate times in the interval $[0, t]$. However, as shown in the Supplementary Text, we have $P(\gamma_{\{0,\tau \}} | \gamma_{[0,t]}) = P(\gamma_{\{0,\tau \}}|\gamma_{\{0, t \}})$, and then $\langle O(t) | \gamma_{[0,t]} \rangle = \langle O(t) | \gamma_{\{0,t\}} \rangle$.

We identify the following martingale process (for a proof see Sec.~\ref{sec:proofs})
\begin{equation}\label{eq:mar}
  \langle e^{-\Delta S_\mathrm{mar}(\tau) - \delta_\mathrm{q}(\tau)} | \gamma_{[0,t]} \rangle = e^{-\Delta S_\mathrm{mar}(t) - \mathcal{\delta}_\mathrm{q}(t)},
\end{equation}
where we recall the definition of the quantum version of the stochastic distinguishability
\begin{equation}\label{eq:dq}
\delta_\mathrm{q}(t) = \ln \left( \frac{\langle \psi(t) | \rho(t) |\psi(t) \rangle}{\langle \psi(t) | \Theta^\dagger \tilde{\rho}(\tau - t) \Theta |\psi(t) \rangle} \right). 
\end{equation}
 Notably, the average of $\delta_\mathrm{q}(t)$ at fixed times $t \leq \tau$ equals the relative entropy (Kullback-Leibler divergence) between the forward and backward density operators $\langle \delta_\mathrm{q} (t) \rangle = \sum_\gamma P(\gamma_{[0,t]}) \delta_\mathrm{q}(t) = D[\rho(t) || \Theta^\dagger \tilde{\rho}(t) \Theta] \equiv \mathrm{Tr}[\rho(t) (\ln \rho(t) - \ln \Theta^\dagger \tilde{\rho}(t) \Theta)]$, which provides an information-theoretical measure of the irreversibility in the process~\cite{STR1,SsagawaREV}.
 Moreover, we proof in Sec.~\ref{sec:proofs} that the uncertainty entropy production in Eq.~(5) of the main text fulfills the generalized fluctuation relation 
 \begin{equation} \label{eq:gft}
   \langle e^{-\Delta S_\mathrm{unc}(\tau)} | \gamma_{[0,t]} \rangle = 1.
 \end{equation}
 
Applying Doob's optional sampling theorem~\cite{SDoob} to the martingale process in Eq.~\eqref{eq:mar}, and using the expression of the split~\eqref{eq:split} of entropy production, we obtain 
\begin{equation}\label{eq:stopmar}
\langle e^{-\Delta S_\mathrm{tot} - \delta_\mathrm{q} + \Delta S_\mathrm{unc}} \rangle_\mathcal{T} = 1,
\end{equation}
with $\Delta S_\mathrm{tot}$ given in Eq.~\eqref{eq:ep} and the average $\langle O \rangle_\mathcal{T} = \sum_\gamma P (\gamma_{\{0, \mathcal{T}\}}) O(\mathcal{T})$ is taken over stopped trajectories.
Here $\mathcal{T}$ is a bounded stopping time, meaning that $\mathcal{T} < c$ for some arbitrary constant $c$. A proof of Eq.~\eqref{eq:stopmar} is given in Sec.~\ref{sec:proofs}. If we assume a single thermal reservoir, hence we get Eq.~(4) in the main text in terms of the work by means of Eq.~\eqref{eq:epw}.

\subsection{Classical Martingale theory} \label{sub:martinc}
The classical limit of our results is obtained when the whole evolution occurs in the Hamiltonian eigenbasis, $[\rho(t), H(t)] = 0$ at all times. Then the stochastic wavefunction $|\psi(t)\rangle$ is always an eigenstate of $\rho(t)$, that is $\langle \psi(t) | \rho(t) |\psi(t)\rangle = \langle n(t) | \rho(t) |n(t)\rangle = p_{n(t)}(t) \equiv \varrho(n(t),t)$. This leads to classical trajectories where every jump corresponds to a change in the system micro-state and therefore we get $\gamma_{\{ 0, \tau\}} = \{n(t) \}_{t=0}^{\tau}$, while the initial and final measurements of the two-measurements scheme become superfluous.

Therefore we recover from Eq.~\eqref{eq:smar} the classical expression of the stochastic entropy production~\cite{SSeifertREV}, namely
\begin{equation}
 \Delta S_\mathrm{mar}(t) = \Delta S_\mathrm{tot}(t) = \Delta S(t) - \beta Q(t).
\end{equation}
Analogously, from Eq.~(5) of the main text we obtain $\Delta S_\mathrm{unc}(t) = 0$ for all $t$ and Eq.~\eqref{eq:dq} reduces to its classical counterpart in Eq.~(2) of the main text. Substituting into Eq.~\eqref{eq:mar} we obtain the Martingale:
\begin{equation}\label{eq:marc}
  \langle e^{-\Delta S_\mathrm{tot}(\tau) - \delta(\tau)} | \gamma_{[0,t]} \rangle = e^{-\Delta S_\mathrm{tot}(t) - \mathcal{\delta}(t)},
\end{equation}
leading to the following stopping-times fluctuation theorem for the entropy production, and second law at stopping times:
\begin{equation}\label{eq:stopmarc}
\langle e^{-\Delta S_\mathrm{tot} - \delta} \rangle_\mathcal{T} = 1 ~~;~~ \langle \Delta S_\mathrm{tot} \rangle_\mathcal{T} \geq -\langle \delta \rangle_\mathcal{T}.
\end{equation}
Finally, using the expression for the entropy production in Eq.~\eqref{eq:epw} in terms of work and free energy, we obtain the second law inequality in Eq.~(1) of the main text and the work fluctuation relation in Eq.~(3). If the system remains in a (time-symmetric) steady state during the evolution, that is, $\tilde{\varrho}(n,\tau-t) = \varrho(n,t) \equiv \varrho_{\rm st}(n)$, then $\delta(t) = 0$ for all $t$, and our results reduce to the steady-state second law at stopping times, $\langle \Delta S_\mathrm{tot} \rangle_{\mathcal{T}} \geq 0$, or equivalently $\langle W \rangle_{\mathcal{T}}  - \langle \Delta F \rangle_{\mathcal{T}} \geq  0$~\cite{Sneri2}.

\section{Proofs of main fluctuation relations} \label{sec:proofs}
Here we provide the proof of the main fluctuation relations leading to our classical and quantum martingale theory for driven systems in arbitrary out-of-equilibrium states. We first provide a direct proof of the martingality of the classical process $e^{-\Delta S_\mathrm{tot}(t) - \delta(t)}$, and the classical stopping-times work fluctuation fluctuation relation in Eq.~(3) of the main text. Then we proof our quantum results in full generality, namely, the martingality of the process $e^{-\Delta S_\mathrm{mar}(t) - \delta_\mathrm{q}(t)}$ as stated in Eq.~\eqref{eq:mar}, where $\Delta S_\mathrm{mar}(t)$ is the martingale entropy production as introduced in Eq.~\eqref{eq:smar} and $\delta_\mathrm{q}(t)$ is the stochastic distinguishability in Eq.~\eqref{eq:dq}. As a second step we proof the generalized fluctuation theorem $\langle e^{-\Delta S_\mathrm{unc}(\tau)} | \gamma_{[0,t]} \rangle$ introduced in Sec.~\ref{sub:martinq}. Finally, we provide a proof of the stopping-time work fluctuation relation in Eq.~\eqref{eq:stopmar}, from which all other results directly follow, including the classical results and the quantum fluctuation relation in Eq.~(4) of the main text.

\subsection{Classical proofs}
{\bf \flushleft Proof of Martingality in Eq.~\eqref{eq:marc}.} We provide a proof of Eq.~\eqref{eq:marc} in Sec.~\ref{sub:martinc}, whose main passages are explained inline below:
\begin{align} 
 \langle e^{-\Delta S_\mathrm{tot}(\tau) - \delta(\tau)} | x_{[0,t]} \rangle &\equiv  \sum_{x_{[t,\tau]}} e^{-\Delta S_\mathrm{tot}(\tau) - \delta(\tau)} P(x_{[0,\tau]} | x_{[0,t]}) = \sum_{x_{[t,\tau]}} e^{-\Delta S_\mathrm{tot}(\tau)} \frac{{P}({x}_{[0,\tau]})}{P(x_{[0,t]})} \label{proof1-1} \\
 &= e^{-\Delta S_\mathrm{tot}(t)} \sum_{x_{[t, \tau]}} \frac{\tilde{P}(\tilde{x}_{[0,\tau]})}{\tilde{P}(\tilde{x}_{[0,t]})} =  e^{-\Delta S_\mathrm{tot}(t)} \frac{\tilde{\rho}(\tilde{x}(t), \tau-t) \tilde{P}[\tilde{x}(0) | \tilde{x}(t)]}{\tilde{P}(\tilde{x}_{[0,t]})} \label{proof1-2} \\
 &=e^{-\Delta S_\mathrm{tot}(t)} \frac{\tilde{\rho}(\tilde{x}(t), \tau-t)}{\rho(\tilde{x}(t),t)} = e^{-\Delta S_\mathrm{tot}(t) - \delta(t)}. \qquad \square \label{proof1-3}
\end{align}
In Eq.~\eqref{proof1-1} we used Bayes' theorem for the conditional probability $P(x_{[0,\tau]} | x_{[0,t]}) = P(x_{[0,t]} | x_{[0,\tau]}) {P}({x}_{[0,\tau]})/P(x_{[0,t]})$ with $P(x_{[0,t]} | x_{[0,\tau]}) = 1$ since $t \leq \tau$, and the fact that $\delta(\tau) = 0$ from our choice of the time-reversed process, i.e. the initial state of the time-reversed process is the final state of the forward one.
In the first equality of~\eqref{proof1-2} we used the explicit form of the stochastic entropy production up to the final time $\tau$, that is, $e^{-\Delta S_\mathrm{tot}(\tau)} = \tilde{P}(\tilde{x}_{[0,\tau]})/P(x_{[0, \tau]})$, and identified $e^{-\Delta S_\mathrm{tot}(t)} = \tilde{P}(\tilde{x}_{[0,t]})/P(x_{[0,t]})$ as the entropy production up to time $t$. In the second equality of~\eqref{proof1-2} we performed the sum over trajectories $x_{[t, \tau]}$, leading to the marginalization:
\begin{equation}\label{eq:marginalizationc}
 \sum_{x_{[t, \tau]}} \tilde{P}(\tilde{x}_{[0,\tau]}) = \sum_{x_{[t, \tau]}} \rho(x(\tau), \tau) \tilde{P}[x(0) | x(\tau)] = \tilde{\rho}(x(t), \tau-t) \tilde{P}[x(0) | x(t)],
\end{equation}
where $\tilde{P}[x(0) | x(t)]$ is the probability for reaching $x(0)$ from $x(t)$ in the time-reversed process, and we used Markovianity in order to reach the final equality in~\eqref{eq:marginalizationc}. Notice that here we are assuming $x$ an even variable under time-reversal, but the proof can be straightforwardly extended to odd variables (see quantum proofs below). Subsequently, we use in Eq.~\eqref{proof1-3} the explicit expression of the path probability $\tilde{P}(x_{[0,t]}) = \rho(x(t), t) \tilde{P}[x(0) | x(t)]$. Finally, we identify the expression for the stochastic distinguishability $\delta (t) = \ln [\rho(x(t),t)/\tilde{\rho}(x(t),\tau - t)]$ as follows from Eq.~(2) of the main text, which completes the proof.  
    
{\bf \flushleft Proof of the stopping-time work fluctuation relation, Eq.~(3) in the main text.} The stopping-time work fluctuation relation follows from Doob's optional stopping theorem~\cite{SDoob}, which holds for generic Martingale processes $M(t)$. Let $\mathcal{T}$ be a bounded stopping time, i.e. $\mathcal{T} < c$ for some arbitrary constant $c$. Doob's optional stopping theorem states that $\langle M \rangle_\T = \langle M(\tau) \rangle = \langle M(0) \rangle$ for any stopping time obeying $\mathcal{T} \leq \tau$~\cite{SWilliams}.

Identifying as the Martingale process $M(t)=e^{-\Delta S_\mathrm{tot}(t) - \delta(t)}$, we obtain:
\begin{equation}
 \langle e^{- \Delta S_\mathrm{tot} - \delta} \rangle_\T = \langle e^{- \Delta S_\mathrm{tot}(\tau) - \delta(\tau)} \rangle = 1, \qquad \square
\end{equation}
where the last equality follows by noticing that $\langle M(0) \rangle = 1$, since $\Delta S_\mathrm{tot}(0) = \delta(0) = 0$. Note that the stopping times $\mathcal{T}$ need to occur in the interval $[0,\tau]$ for any arbitrary finite $\tau$. However, following \cite{SWilliams}, we may also take $\tau \rightarrow \infty$ whenever $|M(t)|$ is finite for all $t \equiv \mathrm{min}(\mathcal{T}, \tau)$. 

Using the expression for the stochastic entropy production in terms of the work, $\Delta S_\mathrm{tot}(t) = \beta[W(t) - \Delta F(t)]$ [see Eq.~\eqref{eq:epw}], we obtain the stopping-time work fluctuation relation in Eq.~(3) of the main text. Finally, the generalized second law inequality at stopping times follows by applying Jensen's inequality to the above equation, that is, $e^{\langle -\Delta S_\mathrm{tot} - \delta \rangle_\T} \leq \langle e^{-\Delta S_\mathrm{tot} - \delta} \rangle_\T = 1$, which implies $\langle \Delta S_\mathrm{tot}\rangle_\T + \langle \delta \rangle_\T \geq 0$. Again substituting $\Delta S_\mathrm{tot}(t) = \beta[W(t) - \Delta F(t)]$ we recover inequality~(1) of the main text.

\subsection{Quantum proofs}
Before going into the quantum proofs it is convenient to first recall some of the properties of trajectory probabilities in the context of quantum jumps. We denote as $P(\gammatau)$ the probability of a trajectory $\gammatau = \{n(0); \R_0^\tau; n(\tau)\}$ associated to the implementation of the protocol $\Lambda$, and starting in an eigenstate $\ket{n(0)}$ of the initial state $\rho(0)$ with corresponding eigenvalue $p_{n(0)}(0)$. According to Born's rule, this probability can be written as $P(\gammatau) = P[n(\tau); \R_0^\tau | n(0)]~ p_{n(0)}(0)$, where the conditional probability reads $P[n(\tau); \R_0^\tau | n(0)] = |\bra{n(\tau)} \Li_0^\tau \ket{n(0)}|^2$, and $\rho(\tau) = \sum_{n(\tau)} p_{n(\tau)}(\tau)$ $\ket{n(\tau)} \bra{n(\tau)}$ is the spectral decomposition of the average state of the system $\rho(\tau)$ at the final time $\tau$. Here we introduced $\Li_0^\tau$ as the operator generating the normalized wavefunction 
\begin{equation}\label{eq:top}
\ket{\psi(\tau)} = \frac{\Li_0^\tau \ket{n(0)}}{\sqrt{\langle \Li_0^{\tau \dagger} \Li_0^\tau \rangle_{n(0)}}}, 
\end{equation}
corresponding to the environmental record $\R_0^\tau$, which verifies the stochastic Sch\"odinger equation~\eqref{eq:stochastic}. Using Eq.~\eqref{eq:top} we can rewrite the conditional probability of a trajectory as 
\begin{equation}\label{eq:c}
 P[n(\tau); \R_0^\tau | n(0)] = |\langle n(\tau)| \psi(\tau) \rangle|^2 \langle \Li_0^{\tau \dagger} \Li_0^\tau \rangle_{n(0)}.
\end{equation}
Analogously, the probability of the time-reversed trajectory associated to the time-reversed protocol $\tilde{\Lambda}$ is denoted as $\tilde P (\tilde{\gamma}_{\{0,\tau\}})$, which starts in $\Theta \ket{n(\tau)}$, with $\Theta$ the anti-unitary time-reversal operator. Since we have chosen the initial state of the time-reversed process to equal the (inverted) final state of the forward one we have $\tilde{P}(\tilde{\gamma}_{\{0,\tau\}}) = \tilde{P}[n(0) \tilde{\R}_\tau^0 | n(\tau)]~ p^\tau_{n(\tau)}$, where the conditional probability reads $\tilde{P}[n(0) \tilde{\R}_\tau^0 | n(\tau)] = |\bra{n(0)} \Theta^\dagger \tilde{\mathcal{L}}_\tau^0 \Theta \ket{n(\tau)}|^2$ with $\tilde{\Li}_\tau^0$ the corresponding operator generating the backward evolution associated to the time-reversed record $\tilde{\mathcal{R}}_\tau^0$. 

Remarkably, the conditional probabilities for forward and time-reversed trajectories obey the following detailed-balance relation:
\begin{equation} \label{eq:db}
 P[n(\tau); \R_0^\tau | n(0)] = \tilde{P}[n(0) \tilde{\R}_\tau^0 | n(\tau)] ~e^{\Delta S_\mathrm{env}(\R_0^\tau)},
\end{equation}
where we denoted $\Delta S_\mathrm{env}(\R_0^\tau) = \sum_{j=1}^J \Delta S_\mathrm{env}^{k_j}$ as the total entropy change in the environment along the trajectory $\gammatau$ associated with the jumps in the environmental measurement record $\R_0^\tau = \{(k_1, t_1), ..., (k_J, t_J) \}$. The relation Eq.~\eqref{eq:db} follows from the relation between forward and time-reverse trajectory generators 
\begin{equation}\label{eq:dbo}
\Theta^\dagger \tilde{\mathcal{L}}_\tau^0 \Theta = \Li_0^{\tau \dagger}~\exp({- \Delta S_\mathrm{env}(\R_0^\tau)/2}), 
\end{equation}
generalizing micro-reversibility to open quantum systems~\cite{Smanzano}.

Finally, it is helpful to stress some properties of the conditional probability $P(\gamma_{\{0,\tau\}} | \gamma_{[0,t]}) \equiv P(\gamma_{\{0,t\}}, \gamma_{[0,\tau]})/P(\gamma_{[0,\tau]})$, where $\gamma_{[0,t]} \equiv \bigcup_{s=0}^t \gamma_{\{0,s\}}$ denotes the ensemble of trajectories eventually stopped at all intermediate times in the interval $[0,t]$. This conditional probability fulfills: 
\begin{align}\label{cond1}
 P(\gamma_{\{0,\tau\}} | \gamma_{[0,t]}) &= \frac{P(\gamma_{\{0,\tau\}})}{P(\gamma_{[0,t]})} P(\gamma_{[0,t]}|\gamma_{\{0,\tau\}}) = \frac{P(\gamma_{\{0,\tau\}})}{P(\gamma_{\{0,t\}})} P(\gamma_{\{0,t\}}|\gamma_{\{0,\tau\}}) \\ \label{cond2}
 &= P(\gamma_{\{0,\tau\}} | \gamma_{\{0,t\}}).
 \end{align} 
Here in the first equality of Eq.~\eqref{cond1} we used Bayes' theorem, $P(\gamma_{\{0,t\}}, \gamma_{[0,\tau]}) = P(\gamma_{\{0,t\}})P(\gamma_{[0,t]}|\gamma_{\{0, \tau\}})$. In the second equality we used that the probabilities of virtual measurements at intermediate times in $\gamma_{[0,t]}$ are independent, which implies $P(\gamma_{[0,t]}) = P(\gamma_{\{0,t\}}) \prod_{s=0}^t |\langle n(s)| \psi(s) \rangle|^2$ and $P(\gamma_{[0,t]}|\gamma_{\{0,\tau\}}) = P(\gamma_{\{0,t\}}|\gamma_{\{0, \tau\}}) \prod_{s=0}^t|\langle n(s)| \psi(s) \rangle|^2$, where in both cases $|\langle n(s)| \psi(s) \rangle|^2$ is the probability that the stochastic wavefunction $\ket{\psi (s)}$ following a trajectory $\gamma_{\{0,\tau\}}$ is found to be in the eigenstate $\ket{n(s)}$ of $\rho(s)$ at time $s$. Finally, to reach Eq.~\eqref{cond2} we used again Bayes' rule to swap back conditions. Equation~\eqref{cond2} implies that conditional averages of arbitrary stochastic functionals along trajectories with respect to ensembles $\gamma_{[0,t]}$ are equivalent to conditional averages with respect to single trajectories $\gamma_{[0,t]}$~\cite{Sours}.

{\bf \flushleft Proof of Martingality in Eq.~\eqref{eq:mar}.} We now proceed with the proof of Eq.~\eqref{eq:mar} in Sec.~\ref{sub:martinq}, whose main passages are explained inline below:
\begin{align} 
 \langle e^{-\Delta S_\mathrm{mar}(\tau) - \delta_\mathrm{q}(\tau)} | \gamma_{[0,t]} \rangle &\equiv  \sum_{n(\tau)} \sum_{\R_t^\tau} e^{-\Delta S_\mathrm{mar}(\tau) - \delta_\mathrm{q}(\tau)} P(\gammatau | \gamma_{[0,t]}) \nonumber \\ \label{proof3-1}
 &= \sum_{n(\tau)} \sum_{\R_t^\tau} e^{-\Delta S_\mathrm{tot}(\tau) + \Delta S_\mathrm{unc}(\tau) - \delta_\mathrm{q}(\tau)} \frac{P(\gamma_{\{0,\tau\}})}{P(\gamma_{\{0,t\}})} |\langle n(t)| \psi(t) \rangle|^2 \\   \label{proof3-2}
 &= \sum_{n(\tau)} \sum_{\R_t^\tau} e^{\Delta S_\mathrm{unc}(\tau)} \frac{\tilde{P}(\tilde{\gamma}_{\{0,\tau\}})}{P(\gamma_{\{0,t\}})} |\langle n(t)| \psi(t) \rangle|^2 \\ \label{proof3-3}
 &= e^{-\Delta S_\mathrm{tot}(t)} \sum_{n(\tau)} \sum_{\R_t^\tau} e^{\Delta S_\mathrm{unc}(\tau)} \frac{\tilde{P}(\tilde{\gamma}_{\{0,\tau\}})}{\tilde{P}(\tilde{\gamma}_{\{0,t\}})} |\langle n(t)| \psi(t) \rangle|^2 \\ \label{proof3-4}
 &= e^{-\Delta S_\mathrm{tot}(t)} \sum_{n(\tau)} \sum_{\R_t^\tau} \langle \psi(\tau) | \rho(\tau) | \psi(\tau) \rangle \frac{\tilde{P}[n(0);\tilde{\mathcal{R}}_\tau^0 | n(\tau)]}{\tilde{P}(\tilde{\gamma}_{\{0,t\}})} |\langle n(t)| \psi(t) \rangle|^2 \\ \label{proof3-5}  
 &= e^{-\Delta S_\mathrm{tot}(t)} \sum_{n(\tau)} \sum_{\R_t^\tau} \sum_{k(\tau)} \frac{p_{k(\tau)}(\tau) P[k(\tau); \R_0^\tau|n(0)]}{\langle \Li_0^{\tau \dagger} \Li_0^\tau  \rangle_{n(0)}}  \frac{\tilde{P}[n(0); \tilde \R_\tau^0|n(\tau)]}{\tilde{P}(\tilde{\gamma}_{\{0,t\}})}  |\langle n(t)| \psi(t) \rangle|^2  \\ \label{proof3-6}
 &= e^{-\Delta S_\mathrm{tot}(t)} \sum_{k(\tau)} \sum_{\R_t^\tau} \frac{p_{k(\tau)}(\tau) \tilde P[n(0); \tilde \R_\tau^0|k(\tau)]}{\tilde{P}(\tilde{\gamma}_{\{0,t\}})}  |\langle n(t)| \psi(t) \rangle|^2  \\ \label{proof3-7}
 &= e^{-\Delta S_\mathrm{tot}(t)} \frac{\bra{\psi(t)} \Theta^\dagger \tilde{\rho}(\tau - t) \Theta \ket{\psi(t)} e^{-\Delta S_\mathrm{env}(\R_0^t)} \langle \Li_0^{t \dagger} \Li_0^t \rangle_{n(0)}}{\tilde{P}(\tilde{\gamma}_{\{0,t\}})}  |\langle n(t)| \psi(t) \rangle|^2 \\ \label{proof3-8}
&= e^{-\Delta S_\mathrm{tot}(t)} \frac{\bra{\psi(t)} \Theta^\dagger \tilde{\rho}(\tau - t) \Theta \ket{\psi(t)}}{p_{n(t)}(t)} \\ \label{proof3-9}
&= e^{-\Delta S_\mathrm{tot}(t)} \frac{\bra{\psi(t)} \Theta^\dagger \tilde{\rho}(\tau - t) \Theta \ket{\psi(t)}}{\langle \psi(t) | \rho(t) \psi(t) \rangle} \frac{\langle \psi(t) | \rho(t) \psi(t) \rangle}{p_{n(t)}(t)} \\ \label{proof3-10} 
 &= e^{-\Delta S_\mathrm{tot}(t) + \Delta S_\mathrm{unc}(t) - \delta_\mathrm{q}(t)} = e^{-\Delta S_\mathrm{mar}(t) - \delta_\mathrm{q}(t)}. \qquad \square
\end{align}
In the second line \eqref{proof3-1} we used $\Delta S_\mathrm{mar}(\tau) = \Delta S_\mathrm{tot}(\tau) - \Delta S_\mathrm{unc}(\tau)$, together with Eq.~\eqref{cond1} and introduced the explicit expression of the conditional probability $P(\gamma_{\{0,t\}}|\gamma_{\{0, \tau\}}) = |\langle n(t)| \psi(t) \rangle|^2$. We subsequently used that $\delta_\mathrm{q}(\tau)=0$, as follows from the fact that the initial state of the time-reversed process is the final state of the forward one. Then using $e^{-\Delta S_\mathrm{tot}(\tau)} = \tilde{P}(\tilde{\gamma}_{\{0,\tau\}}) / P(\gamma_{\{0,\tau\}})$, as follows from Eq.~(9) in the Methods, we reach the third line \eqref{proof3-2}. In line \eqref{proof3-3} we introduced $e^{-\Delta S_\mathrm{tot}(t)} = \tilde{P}(\tilde{\gamma}_{\{0,t\}}) / P(\gamma_{\{0,t\}})$. 
In \eqref{proof3-4} we substituted the definition of the uncertainty entropy production $\Delta S_\mathrm{unc}(\tau)$ in Eq.~(5) of the main text and expanded the probability of the backward trajectory $\tilde{P}(\tilde{\gamma}_{\{0,\tau\}}) = p_{n(\tau)}(\tau) \tilde{P}[n(0);\tilde{\mathcal{R}}_\tau^0 | n(\tau)]$. Equation  \eqref{proof3-5} is obtained after introducing the spectral decomposition of $\rho(\tau)$ to get $\langle \psi(\tau) | \rho(\tau) | \psi(\tau) \rangle = \sum_{k(\tau)} p_{k(\tau)}(\tau)|\langle \psi(\tau) | k(\tau) \rangle|^2$ and then using Eq.~\eqref{eq:c}. We then used the detailed-balance relation in Eq.~\eqref{eq:db} to both 
$P[k(\tau); \R_0^\tau|n(0)]$ and $\tilde{P}[n(0); \tilde \R_\tau^0|n(\tau)]$ and perform the sum over $n(\tau)$, such that $\sum_{n(\tau)} P[n(\tau); \R_0^\tau|n(0)] = \langle \Li_0^{\tau \dagger} \Li_0^\tau  \rangle_{n(0)}$ to reach Eq.~\eqref{proof3-6}. Summing \eqref{proof3-6} over $k(\tau)$ and the measurement record $\R_t^\tau$ leads to the marginalization: 
\begin{equation} \label{eq:marginalization}
\sum_{k(\tau)} \sum_{\R_t^\tau}  p_{k(\tau)}(\tau) \tilde{P}[n(0); \tilde\R_\tau^0 | k(\tau)]  = \bra{n(0)} \Theta^\dagger \tilde{\Li}_\tau^0 ~ \tilde{\rho}(\tau - t) \tilde{\Li}_\tau^{0 \dagger} \Theta \ket{n(0)}, 
\end{equation}
where $\tilde{\rho}(\tau -t)$ is the density operator generated in the time-reversed dynamics under the protocol $\tilde{\Lambda}$. Upon using operator micro-reversibility in Eq.~\eqref{eq:dbo} together with the definition of the stochastic wavefunction in Eq.~\eqref{eq:top}, Eq.~\eqref{eq:marginalization} leads to Eq.~\eqref{proof3-7}. Now expanding $\tilde{P}(\tilde{\gamma_{\{0, t\}}}) = p_{n(t)}(t) \tilde{P}[n(0);\tilde{\R}_t^0 | n(t)]$ in the denominator and noticing that $e^{-\Delta S_\mathrm{env}(\R_0^t)} \langle \Li_0^{t \dagger} \Li_0^t \rangle_{n(0)} |\langle n(t)| \psi(t) \rangle|^2 = \tilde{P}[n(0);\tilde{\R}_t^0 | n(t)]$ in the numerator [as follows by combining Eqs.~\eqref{eq:c} and \eqref{eq:db}] we reach Eq.~\eqref{proof3-8}. Finally, multiplying and dividing by $\langle \psi(t) | \rho(t) \psi(t) \rangle$ we get Eq.~\eqref{proof3-9}, which upon identifying the terms $\delta_\mathrm{q}(t)$ and $\Delta S_\mathrm{unc}(t)$, leads to the final line \eqref{proof3-10}.

Since Eqs.~\eqref{proof3-1}-\eqref{proof3-10} are verified and $e^{-\Delta S_\mathrm{mar}(t)} < \infty$ is bounded, we conclude that $\Delta S_\mathrm{mar} - \delta_\mathrm{q}$ is an exponential martingale. Choosing $t = 0$ we recover from Eq.~\eqref{proof3-10} the integral fluctuation theorem $\langle  e^{-\Delta S_\mathrm{mar}(\tau) - \delta_\mathrm{q}(\tau)} \rangle = 1$. 

{\bf \flushleft Proof of the generalized fluctuation theorem, Eq.~\eqref{eq:gft}.} As stated in Sec.~\ref{sub:martinq}, we obtain the following fluctuation theorem for the uncertainty entropy production $\Delta S_\mathrm{unc}(t)$ [Eq.(5) of the main text]:
\begin{align}
  \langle e^{-\Delta S_\mathrm{unc}(\tau)} | \gamma_{[0,t]} \rangle &\equiv  \sum_{n(\tau)} \sum_{\R_t^\tau} e^{-\Delta S_\mathrm{unc}(\tau)} P(\gammatau | \gamma_{[0,t]}) \nonumber = \sum_{n(\tau)} \sum_{\R_t^\tau} e^{-\Delta S_\mathrm{unc}(\tau)} \frac{P(\gamma_{\{0,\tau\}})}{P(\gamma_{\{0,t\}})} |\langle n(t)| \psi(t) \rangle|^2, \\
  &= \sum_{n(\tau)} \sum_{\R_t^\tau} \frac{p_{n(\tau)}(\tau)}{\langle \psi(\tau)| \rho(\tau)| \psi(\tau) \rangle} \frac{P(\gamma_{\{0,\tau\}})}{P(\gamma_{\{0,t\}})} |\langle n(t)| \psi(t) \rangle|^2 \label{line1} \\ 
  &= \sum_{n(\tau)} \sum_{\R_t^\tau} \frac{p_{n(\tau)(\tau)}}{\langle \psi(\tau)| \rho(\tau)| \psi(\tau) \rangle} \frac{ \langle \Li_0^{\tau \dagger} \Li_0^\tau \rangle_{n(0)}}{  \langle \Li_0^{t \dagger} \Li_0^t \rangle_{n(0)}} |\langle n(\tau)| \psi(t) \rangle|^2 \label{line2} \\
&= \sum_{\R_t^\tau} \frac{\langle \Li_0^{\tau \dagger} \Li_0^\tau \rangle_{n(0)}}{\langle \Li_0^{t \dagger} \Li_0^t \rangle_{n(0)}} = 1, \qquad \square \label{line3}
  \end{align}
where in the first line we introduced the expression of the conditional probability and in \eqref{line1} the one of $\Delta S_\mathrm{unc}(\tau)$. Equation~\eqref{line2} follows by introducing the expressions of the trajectory probabilities $P(\gammatau)$ and $P(\gammat)$, using Eq.~\eqref{eq:c} and cancelling terms. Noticing that $\sum_{n(\tau)}p_{n(\tau)(\tau)} |\langle n(\tau)| \psi(t) \rangle|^2 = \langle \psi(\tau)| \rho(\tau)| \psi(\tau) \rangle$ we arrive to Eq.~\eqref{line3}, which upon summing the numerator over the environmental record $\R_t^\tau$, gives the final result $1$.

Finally, we notice that whenever the state of the system becomes symmetric under time-reversal, we have $\rho(t) = \Theta^\dagger \tilde{\rho}(\tau - t) \Theta \equiv \rho_\mathrm{ss}$ for all $t \in [0, \tau]$, and therefore $\delta_\mathrm{q} = 0$. In such case we recover the quantum martingale theory for nonequilibrium steady states derived in Ref.~\cite{Sours}.

{\bf \flushleft Proof of the stopping-time fluctuation relation, Eq.~\eqref{eq:stopmar}.} As in the classical case above, the stopping-time work fluctuation relation in Eq.~\eqref{eq:stopmar} follows from Doob's optional stopping theorem~\cite{SDoob}, $\langle M \rangle_\T = \langle M(\tau) \rangle = \langle M(0) \rangle$, while in this case we apply it to the quantum Martingale $M(t)=e^{-\Delta S_\mathrm{mar}(t) - \delta_\mathrm{q}(t)}$.

Assuming $\mathcal{T}$ a bounded stopping time obeying $\mathcal{T} \leq \tau$, we have:
\begin{equation}
 \langle e^{- \Delta S_\mathrm{mar} - \delta_\mathrm{q}} \rangle_\T = \langle e^{- \Delta S_\mathrm{mar}(\tau) - \delta_\mathrm{q}(\tau)} \rangle = 1, \qquad \square
\end{equation}
where the last equality follows from $\Delta S_\mathrm{mar}(0) = \delta_\mathrm{q}(0) = 0$. As pointed before, this theorem is also valid for $\tau \rightarrow \infty$ whenever $|M(t)|$ is finite for all $t \equiv \mathrm{min}(\mathcal{T}, \tau)$~\cite{SWilliams}. Finally, the generalized second law inequality at stopping times, follows by applying Jensen's inequality to the above equation, that is, $e^{\langle -\Delta S_\mathrm{mar} - \delta_\mathrm{q} \rangle_\T} \leq \langle e^{-\Delta S_\mathrm{mar} - \delta_\mathrm{q}} \rangle_\T = 1$, which implies $\langle \Delta S_\mathrm{mar}\rangle_\T + \langle \delta_\mathrm{q} \rangle_\T \geq 0$.


\begin{thebibliography}{57}%
\makeatletter
\providecommand \@ifxundefined [1]{%
 \@ifx{#1\undefined}
}%
\providecommand \@ifnum [1]{%
 \ifnum #1\expandafter \@firstoftwo
 \else \expandafter \@secondoftwo
 \fi
}%
\providecommand \@ifx [1]{%
 \ifx #1\expandafter \@firstoftwo
 \else \expandafter \@secondoftwo
 \fi
}%
\providecommand \natexlab [1]{#1}%
\providecommand \enquote  [1]{``#1''}%
\providecommand \bibnamefont  [1]{#1}%
\providecommand \bibfnamefont [1]{#1}%
\providecommand \citenamefont [1]{#1}%
\providecommand \href@noop [0]{\@secondoftwo}%
\providecommand \href [0]{\begingroup \@sanitize@url \@href}%
\providecommand \@href[1]{\@@startlink{#1}\@@href}%
\providecommand \@@href[1]{\endgroup#1\@@endlink}%
\providecommand \@sanitize@url [0]{\catcode `\\12\catcode `\$12\catcode
  `\&12\catcode `\#12\catcode `\^12\catcode `\_12\catcode `\%12\relax}%
\providecommand \@@startlink[1]{}%
\providecommand \@@endlink[0]{}%
\providecommand \url  [0]{\begingroup\@sanitize@url \@url }%
\providecommand \@url [1]{\endgroup\@href {#1}{\urlprefix }}%
\providecommand \urlprefix  [0]{URL }%
\providecommand \Eprint [0]{\href }%
\providecommand \doibase [0]{https://doi.org/}%
\providecommand \selectlanguage [0]{\@gobble}%
\providecommand \bibinfo  [0]{\@secondoftwo}%
\providecommand \bibfield  [0]{\@secondoftwo}%
\providecommand \translation [1]{[#1]}%
\providecommand \BibitemOpen [0]{}%
\providecommand \bibitemStop [0]{}%
\providecommand \bibitemNoStop [0]{.\EOS\space}%
\providecommand \EOS [0]{\spacefactor3000\relax}%
\providecommand \BibitemShut  [1]{\csname bibitem#1\endcsname}%
\let\auto@bib@innerbib\@empty
\bibitem [{\citenamefont {Rex}\ and\ \citenamefont {Leff}(2002)}]{Maxwell}%
  \BibitemOpen
  \bibfield  {author} {\bibinfo {author} {\bibfnamefont {A.}~\bibnamefont
  {Rex}}\ and\ \bibinfo {author} {\bibfnamefont {H.~S.}\ \bibnamefont {Leff}},\
  }\href@noop {} {\emph {\bibinfo {title} {Maxwell's demon 2: entropy,
  classical and quantum information, computing}}}\ (\bibinfo  {publisher}
  {Taylor and Francis},\ \bibinfo {year} {2002})\BibitemShut {NoStop}%
\bibitem [{\citenamefont {Maruyama}\ \emph {et~al.}(2009)\citenamefont
  {Maruyama}, \citenamefont {Nori},\ and\ \citenamefont {Vedral}}]{Review1}%
  \BibitemOpen
  \bibfield  {author} {\bibinfo {author} {\bibfnamefont {K.}~\bibnamefont
  {Maruyama}}, \bibinfo {author} {\bibfnamefont {F.}~\bibnamefont {Nori}},\
  and\ \bibinfo {author} {\bibfnamefont {V.}~\bibnamefont {Vedral}},\
  }\bibfield  {title} {\emph {\bibinfo {title} {Colloquium: The physics of
  {M}axwell's demon and information}},\ }\href@noop {} {\bibfield  {journal}
  {\bibinfo  {journal} {Rev. Mod. Phys}\ }\textbf {\bibinfo {volume} {81}},\
  \bibinfo {pages} {1} (\bibinfo {year} {2009})}\BibitemShut {NoStop}%
\bibitem [{\citenamefont {Sagawa}\ and\ \citenamefont {Ueda}(2008)}]{MDlaws1}%
  \BibitemOpen
  \bibfield  {author} {\bibinfo {author} {\bibfnamefont {T.}~\bibnamefont
  {Sagawa}}\ and\ \bibinfo {author} {\bibfnamefont {M.}~\bibnamefont {Ueda}},\
  }\bibfield  {title} {\emph {\bibinfo {title} {Second law of thermodynamics
  with discrete quantum feedback control}},\ }\href@noop {} {\bibfield
  {journal} {\bibinfo  {journal} {Phys. Rev. Lett.}\ }\textbf {\bibinfo
  {volume} {100}},\ \bibinfo {pages} {080403} (\bibinfo {year}
  {2008})}\BibitemShut {NoStop}%
\bibitem [{\citenamefont {Sagawa}\ and\ \citenamefont {Ueda}(2010)}]{MDlaws2}%
  \BibitemOpen
  \bibfield  {author} {\bibinfo {author} {\bibfnamefont {T.}~\bibnamefont
  {Sagawa}}\ and\ \bibinfo {author} {\bibfnamefont {M.}~\bibnamefont {Ueda}},\
  }\bibfield  {title} {\emph {\bibinfo {title} {Generalized jarzynski equality
  under nonequilibrium feedback control}},\ }\href@noop {} {\bibfield
  {journal} {\bibinfo  {journal} {Phys. Rev. Lett.}\ }\textbf {\bibinfo
  {volume} {104}},\ \bibinfo {pages} {090602} (\bibinfo {year}
  {2010})}\BibitemShut {NoStop}%
\bibitem [{\citenamefont {{Del Rio}}\ \emph {et~al.}(2011)\citenamefont {{Del
  Rio}}, \citenamefont {{\AA}berg}, \citenamefont {Renner}, \citenamefont
  {Dahlsten},\ and\ \citenamefont {Vedral}}]{Rio}%
  \BibitemOpen
  \bibfield  {author} {\bibinfo {author} {\bibfnamefont {L.}~\bibnamefont {{Del
  Rio}}}, \bibinfo {author} {\bibfnamefont {J.}~\bibnamefont {{\AA}berg}},
  \bibinfo {author} {\bibfnamefont {R.}~\bibnamefont {Renner}}, \bibinfo
  {author} {\bibfnamefont {O.}~\bibnamefont {Dahlsten}},\ and\ \bibinfo
  {author} {\bibfnamefont {V.}~\bibnamefont {Vedral}},\ }\bibfield  {title}
  {\emph {\bibinfo {title} {The thermodynamic meaning of negative entropy}},\
  }\href@noop {} {\bibfield  {journal} {\bibinfo  {journal} {Nature}\ }\textbf
  {\bibinfo {volume} {474}},\ \bibinfo {pages} {61--63} (\bibinfo {year}
  {2011})}\BibitemShut {NoStop}%
\bibitem [{\citenamefont {Parrondo}\ \emph {et~al.}(2015)\citenamefont
  {Parrondo}, \citenamefont {Horowitz},\ and\ \citenamefont
  {Sagawa}}]{MDlaws5}%
  \BibitemOpen
  \bibfield  {author} {\bibinfo {author} {\bibfnamefont {J.~M.~R.}\
  \bibnamefont {Parrondo}}, \bibinfo {author} {\bibfnamefont {J.~M.}\
  \bibnamefont {Horowitz}},\ and\ \bibinfo {author} {\bibfnamefont
  {T.}~\bibnamefont {Sagawa}},\ }\bibfield  {title} {\emph {\bibinfo {title}
  {Thermodynamics of information}},\ }\href@noop {} {\bibfield  {journal}
  {\bibinfo  {journal} {Nature Phys.}\ }\textbf {\bibinfo {volume} {11}},\
  \bibinfo {pages} {131--139} (\bibinfo {year} {2015})}\BibitemShut {NoStop}%
\bibitem [{\citenamefont {Lutz}\ and\ \citenamefont
  {Ciliberto}(2015)}]{MDexperimentsREV}%
  \BibitemOpen
  \bibfield  {author} {\bibinfo {author} {\bibfnamefont {E.}~\bibnamefont
  {Lutz}}\ and\ \bibinfo {author} {\bibfnamefont {S.}~\bibnamefont
  {Ciliberto}},\ }\bibfield  {title} {\emph {\bibinfo {title} {Information:
  {F}rom {M}axwell{\rq}s demon to {L}andauer{\rq}s eraser}},\ }\href@noop {}
  {\bibfield  {journal} {\bibinfo  {journal} {Phys. Today}\ }\textbf {\bibinfo
  {volume} {68}},\ \bibinfo {pages} {30} (\bibinfo {year} {2015})}\BibitemShut
  {NoStop}%
\bibitem [{\citenamefont {Gavrilov}\ and\ \citenamefont
  {Bechhoefer}(2016)}]{bechhoefer}%
  \BibitemOpen
  \bibfield  {author} {\bibinfo {author} {\bibfnamefont {M.}~\bibnamefont
  {Gavrilov}}\ and\ \bibinfo {author} {\bibfnamefont {J.}~\bibnamefont
  {Bechhoefer}},\ }\bibfield  {title} {\emph {\bibinfo {title} {Erasure without
  work in an asymmetric double-well potential}},\ }\href@noop {} {\bibfield
  {journal} {\bibinfo  {journal} {Phys. Rev. Lett.}\ }\textbf {\bibinfo
  {volume} {117}},\ \bibinfo {pages} {200601} (\bibinfo {year}
  {2016})}\BibitemShut {NoStop}%
\bibitem [{\citenamefont {Ribezzi-Crivellari}\ and\ \citenamefont
  {Ritort}(2019)}]{ritort}%
  \BibitemOpen
  \bibfield  {author} {\bibinfo {author} {\bibfnamefont {M.}~\bibnamefont
  {Ribezzi-Crivellari}}\ and\ \bibinfo {author} {\bibfnamefont
  {F.}~\bibnamefont {Ritort}},\ }\bibfield  {title} {\emph {\bibinfo {title}
  {Large work extraction and the {L}andauer limit in a continuous {M}axwell
  demon}},\ }\href@noop {} {\bibfield  {journal} {\bibinfo  {journal} {Nature
  Phys.}\ }\textbf {\bibinfo {volume} {15}},\ \bibinfo {pages} {660--664}
  (\bibinfo {year} {2019})}\BibitemShut {NoStop}%
\bibitem [{\citenamefont {Camati}\ \emph {et~al.}(2016)\citenamefont {Camati},
  \citenamefont {Peterson}, \citenamefont {Batalhao}, \citenamefont {Micadei},
  \citenamefont {Souza}, \citenamefont {Sarthour}, \citenamefont {Oliveira},\
  and\ \citenamefont {Serra}}]{MDQexperiments2}%
  \BibitemOpen
  \bibfield  {author} {\bibinfo {author} {\bibfnamefont {P.~A.}\ \bibnamefont
  {Camati}}, \bibinfo {author} {\bibfnamefont {J.~P.}\ \bibnamefont
  {Peterson}}, \bibinfo {author} {\bibfnamefont {T.~B.}\ \bibnamefont
  {Batalhao}}, \bibinfo {author} {\bibfnamefont {K.}~\bibnamefont {Micadei}},
  \bibinfo {author} {\bibfnamefont {A.~M.}\ \bibnamefont {Souza}}, \bibinfo
  {author} {\bibfnamefont {R.~S.}\ \bibnamefont {Sarthour}}, \bibinfo {author}
  {\bibfnamefont {I.~S.}\ \bibnamefont {Oliveira}},\ and\ \bibinfo {author}
  {\bibfnamefont {R.~M.}\ \bibnamefont {Serra}},\ }\bibfield  {title} {\emph
  {\bibinfo {title} {{E}xperimental rectification of entropy production by
  {M}axwell's demon in a quantum system}},\ }\href@noop {} {\bibfield
  {journal} {\bibinfo  {journal} {Phys. Rev. Lett.}\ }\textbf {\bibinfo
  {volume} {117}},\ \bibinfo {pages} {240502} (\bibinfo {year}
  {2016})}\BibitemShut {NoStop}%
\bibitem [{\citenamefont {Cottet}\ \emph {et~al.}(2017)\citenamefont {Cottet},
  \citenamefont {Jezouin}, \citenamefont {Bretheau}, \citenamefont
  {Campagne-Ibarcq}, \citenamefont {Ficheux}, \citenamefont {Anders},
  \citenamefont {Auff{\`e}ves}, \citenamefont {Azouit}, \citenamefont
  {Rouchon},\ and\ \citenamefont {Huard}}]{MDQexperiments3}%
  \BibitemOpen
  \bibfield  {author} {\bibinfo {author} {\bibfnamefont {N.}~\bibnamefont
  {Cottet}}, \bibinfo {author} {\bibfnamefont {S.}~\bibnamefont {Jezouin}},
  \bibinfo {author} {\bibfnamefont {L.}~\bibnamefont {Bretheau}}, \bibinfo
  {author} {\bibfnamefont {P.}~\bibnamefont {Campagne-Ibarcq}}, \bibinfo
  {author} {\bibfnamefont {Q.}~\bibnamefont {Ficheux}}, \bibinfo {author}
  {\bibfnamefont {J.}~\bibnamefont {Anders}}, \bibinfo {author} {\bibfnamefont
  {A.}~\bibnamefont {Auff{\`e}ves}}, \bibinfo {author} {\bibfnamefont
  {R.}~\bibnamefont {Azouit}}, \bibinfo {author} {\bibfnamefont
  {P.}~\bibnamefont {Rouchon}},\ and\ \bibinfo {author} {\bibfnamefont
  {B.}~\bibnamefont {Huard}},\ }\bibfield  {title} {\emph {\bibinfo {title}
  {Observing a quantum {M}axwell demon at work}},\ }\href@noop {} {\bibfield
  {journal} {\bibinfo  {journal} {PNAS}\ }\textbf {\bibinfo {volume} {114}},\
  \bibinfo {pages} {7561--7564} (\bibinfo {year} {2017})}\BibitemShut {NoStop}%
\bibitem [{\citenamefont {Williams}(1991)}]{Williams}%
  \BibitemOpen
  \bibfield  {author} {\bibinfo {author} {\bibfnamefont {D.}~\bibnamefont
  {Williams}},\ }\href@noop {} {\emph {\bibinfo {title} {Probability with
  martingales}}}\ (\bibinfo  {publisher} {Cambridge university press},\
  \bibinfo {year} {1991})\BibitemShut {NoStop}%
\bibitem [{\citenamefont {Pliska}(1997)}]{economy}%
  \BibitemOpen
  \bibfield  {author} {\bibinfo {author} {\bibfnamefont {S.}~\bibnamefont
  {Pliska}},\ }\href@noop {} {\emph {\bibinfo {title} {Introduction to
  mathematical finance}}}\ (\bibinfo  {publisher} {Blackwell publishers
  Oxford},\ \bibinfo {year} {1997})\BibitemShut {NoStop}%
\bibitem [{\citenamefont {Chetrite}\ and\ \citenamefont
  {Gupta}(2011)}]{raphaelshamik}%
  \BibitemOpen
  \bibfield  {author} {\bibinfo {author} {\bibfnamefont {R.}~\bibnamefont
  {Chetrite}}\ and\ \bibinfo {author} {\bibfnamefont {S.}~\bibnamefont
  {Gupta}},\ }\bibfield  {title} {\emph {\bibinfo {title} {Two refreshing views
  of fluctuation theorems through kinematics elements and exponential
  martingale}},\ }\href@noop {} {\bibfield  {journal} {\bibinfo  {journal} {J.
  Stat. Phys.}\ }\textbf {\bibinfo {volume} {143}},\ \bibinfo {pages} {543}
  (\bibinfo {year} {2011})}\BibitemShut {NoStop}%
\bibitem [{\citenamefont {Neri}\ \emph {et~al.}(2017)\citenamefont {Neri},
  \citenamefont {Rold{\'a}n},\ and\ \citenamefont {J{\"u}licher}}]{neri}%
  \BibitemOpen
  \bibfield  {author} {\bibinfo {author} {\bibfnamefont {I.}~\bibnamefont
  {Neri}}, \bibinfo {author} {\bibfnamefont {{\'E}.}~\bibnamefont
  {Rold{\'a}n}},\ and\ \bibinfo {author} {\bibfnamefont {F.}~\bibnamefont
  {J{\"u}licher}},\ }\bibfield  {title} {\emph {\bibinfo {title} {Statistics of
  infima and stopping times of entropy production and applications to active
  molecular processes}},\ }\href@noop {} {\bibfield  {journal} {\bibinfo
  {journal} {Phys. Rev. X}\ }\textbf {\bibinfo {volume} {7}},\ \bibinfo {pages}
  {011019} (\bibinfo {year} {2017})}\BibitemShut {NoStop}%
\bibitem [{\citenamefont {Moslonka}\ and\ \citenamefont
  {Sekimoto}(2020)}]{moslonka2020memory}%
  \BibitemOpen
  \bibfield  {author} {\bibinfo {author} {\bibfnamefont {C.}~\bibnamefont
  {Moslonka}}\ and\ \bibinfo {author} {\bibfnamefont {K.}~\bibnamefont
  {Sekimoto}},\ }\bibfield  {title} {\emph {\bibinfo {title} {Memory through a
  hidden martingale process in progressive quenching}},\ }\href@noop {}
  {\bibfield  {journal} {\bibinfo  {journal} {arXiv:2001.04842}\ } (\bibinfo
  {year} {2020})}\BibitemShut {NoStop}%
\bibitem [{\citenamefont {Ge}\ \emph {et~al.}(2018)\citenamefont {Ge},
  \citenamefont {Jia},\ and\ \citenamefont {Jin}}]{ge2018martingale}%
  \BibitemOpen
  \bibfield  {author} {\bibinfo {author} {\bibfnamefont {H.}~\bibnamefont
  {Ge}}, \bibinfo {author} {\bibfnamefont {C.}~\bibnamefont {Jia}},\ and\
  \bibinfo {author} {\bibfnamefont {X.}~\bibnamefont {Jin}},\ }\bibfield
  {title} {\emph {\bibinfo {title} {Martingale structure for general
  thermodynamic functionals of diffusion processes under second-order
  averaging}},\ }\href@noop {} {\bibfield  {journal} {\bibinfo  {journal}
  {arXiv:1811.04529}\ } (\bibinfo {year} {2018})}\BibitemShut {NoStop}%
\bibitem [{\citenamefont {Yang}\ and\ \citenamefont
  {Qian}(2020)}]{yang2020unified}%
  \BibitemOpen
  \bibfield  {author} {\bibinfo {author} {\bibfnamefont {Y.-J.}\ \bibnamefont
  {Yang}}\ and\ \bibinfo {author} {\bibfnamefont {H.}~\bibnamefont {Qian}},\
  }\bibfield  {title} {\emph {\bibinfo {title} {Unified formalism for entropy
  production and fluctuation relations}},\ }\href@noop {} {\bibfield  {journal}
  {\bibinfo  {journal} {Phys. Rev. E}\ }\textbf {\bibinfo {volume} {101}},\
  \bibinfo {pages} {022129} (\bibinfo {year} {2020})}\BibitemShut {NoStop}%
\bibitem [{\citenamefont {Ch{\'e}trite}\ \emph {et~al.}(2019)\citenamefont
  {Ch{\'e}trite}, \citenamefont {Gupta}, \citenamefont {Neri},\ and\
  \citenamefont {Rold{\'a}n}}]{chetrite}%
  \BibitemOpen
  \bibfield  {author} {\bibinfo {author} {\bibfnamefont {R.}~\bibnamefont
  {Ch{\'e}trite}}, \bibinfo {author} {\bibfnamefont {S.}~\bibnamefont {Gupta}},
  \bibinfo {author} {\bibfnamefont {I.}~\bibnamefont {Neri}},\ and\ \bibinfo
  {author} {\bibfnamefont {{\'E}.}~\bibnamefont {Rold{\'a}n}},\ }\bibfield
  {title} {\emph {\bibinfo {title} {Martingale theory for housekeeping heat}},\
  }\href@noop {} {\bibfield  {journal} {\bibinfo  {journal} {EPL}\ }\textbf
  {\bibinfo {volume} {124}},\ \bibinfo {pages} {60006} (\bibinfo {year}
  {2019})}\BibitemShut {NoStop}%
\bibitem [{\citenamefont {Manzano}\ \emph {et~al.}(2019)\citenamefont
  {Manzano}, \citenamefont {Fazio},\ and\ \citenamefont {Rold{\'a}n}}]{ours}%
  \BibitemOpen
  \bibfield  {author} {\bibinfo {author} {\bibfnamefont {G.}~\bibnamefont
  {Manzano}}, \bibinfo {author} {\bibfnamefont {R.}~\bibnamefont {Fazio}},\
  and\ \bibinfo {author} {\bibfnamefont {{\'E}.}~\bibnamefont {Rold{\'a}n}},\
  }\bibfield  {title} {\emph {\bibinfo {title} {Quantum martingale theory and
  entropy production}},\ }\href@noop {} {\bibfield  {journal} {\bibinfo
  {journal} {Phys. Rev. Lett.}\ }\textbf {\bibinfo {volume} {122}},\ \bibinfo
  {pages} {220602} (\bibinfo {year} {2019})}\BibitemShut {NoStop}%
\bibitem [{\citenamefont {Neri}\ \emph {et~al.}(2019)\citenamefont {Neri},
  \citenamefont {Rold{\'a}n}, \citenamefont {Pigolotti},\ and\ \citenamefont
  {J{\"u}licher}}]{neri2}%
  \BibitemOpen
  \bibfield  {author} {\bibinfo {author} {\bibfnamefont {I.}~\bibnamefont
  {Neri}}, \bibinfo {author} {\bibfnamefont {{\'E}.}~\bibnamefont
  {Rold{\'a}n}}, \bibinfo {author} {\bibfnamefont {S.}~\bibnamefont
  {Pigolotti}},\ and\ \bibinfo {author} {\bibfnamefont {F.}~\bibnamefont
  {J{\"u}licher}},\ }\bibfield  {title} {\emph {\bibinfo {title} {Integral
  fluctuation relations for entropy production at stopping times}},\
  }\href@noop {} {\bibfield  {journal} {\bibinfo  {journal} {J. Stat. Mech.}\
  }\textbf {\bibinfo {volume} {2019}},\ \bibinfo {pages} {104006} (\bibinfo
  {year} {2019})}\BibitemShut {NoStop}%
\bibitem [{\citenamefont {Neri}(2020)}]{neriW}%
  \BibitemOpen
  \bibfield  {author} {\bibinfo {author} {\bibfnamefont {I.}~\bibnamefont
  {Neri}},\ }\bibfield  {title} {\emph {\bibinfo {title} {Second law of
  thermodynamics at stopping times}},\ }\href@noop {} {\bibfield  {journal}
  {\bibinfo  {journal} {Phys. Rev. Lett.}\ }\textbf {\bibinfo {volume} {124}},\
  \bibinfo {pages} {040601} (\bibinfo {year} {2020})}\BibitemShut {NoStop}%
\bibitem [{\citenamefont {Sekimoto}(2010)}]{sekimoto}%
  \BibitemOpen
  \bibfield  {author} {\bibinfo {author} {\bibfnamefont {K.}~\bibnamefont
  {Sekimoto}},\ }\href@noop {} {\emph {\bibinfo {title} {Stochastic
  energetics}}},\ Vol.\ \bibinfo {volume} {799}\ (\bibinfo  {publisher}
  {Springer},\ \bibinfo {year} {2010})\BibitemShut {NoStop}%
\bibitem [{\citenamefont {Seifert}(2012)}]{SeifertREV}%
  \BibitemOpen
  \bibfield  {author} {\bibinfo {author} {\bibfnamefont {U.}~\bibnamefont
  {Seifert}},\ }\bibfield  {title} {\emph {\bibinfo {title} {Stochastic
  thermodynamics, fluctuation theorems and molecular machines}},\ }\href@noop
  {} {\bibfield  {journal} {\bibinfo  {journal} {Rep. Prog. Phys.}\ }\textbf
  {\bibinfo {volume} {75}},\ \bibinfo {pages} {126001} (\bibinfo {year}
  {2012})}\BibitemShut {NoStop}%
\bibitem [{\citenamefont {Jarzynski}(2011)}]{JarzynskiREV}%
  \BibitemOpen
  \bibfield  {author} {\bibinfo {author} {\bibfnamefont {C.}~\bibnamefont
  {Jarzynski}},\ }\bibfield  {title} {\emph {\bibinfo {title} {Equalities and
  inequalities: Irreversibility and the second law of thermodynamics at the
  nanoscale}},\ }\href@noop {} {\bibfield  {journal} {\bibinfo  {journal}
  {Annu. Rev. Condens. Matter Phys.}\ }\textbf {\bibinfo {volume} {2}},\
  \bibinfo {pages} {329--351} (\bibinfo {year} {2011})}\BibitemShut {NoStop}%
\bibitem [{\citenamefont {Jarzynski}(1997)}]{jarz}%
  \BibitemOpen
  \bibfield  {author} {\bibinfo {author} {\bibfnamefont {C.}~\bibnamefont
  {Jarzynski}},\ }\bibfield  {title} {\emph {\bibinfo {title} {Nonequilibrium
  equality for free energy differences}},\ }\href@noop {} {\bibfield  {journal}
  {\bibinfo  {journal} {Phys. Rev. Lett.}\ }\textbf {\bibinfo {volume} {78}},\
  \bibinfo {pages} {2690} (\bibinfo {year} {1997})}\BibitemShut {NoStop}%
\bibitem [{\citenamefont {Seifert}(2005)}]{seifert}%
  \BibitemOpen
  \bibfield  {author} {\bibinfo {author} {\bibfnamefont {U.}~\bibnamefont
  {Seifert}},\ }\bibfield  {title} {\emph {\bibinfo {title} {Entropy production
  along a stochastic trajectory and an integral fluctuation theorem}},\
  }\href@noop {} {\bibfield  {journal} {\bibinfo  {journal} {Phys. Rev. Lett.}\
  }\textbf {\bibinfo {volume} {95}},\ \bibinfo {pages} {040602} (\bibinfo
  {year} {2005})}\BibitemShut {NoStop}%
\bibitem [{til()}]{tildenote}%
  \BibitemOpen
  \href@noop {} {}\bibinfo {note} {Here $\tilde{x} = \pm x$ according to the
  parity of the variable $x$ under time reversal.}\BibitemShut {Stop}%
\bibitem [{SM()}]{SM}%
  \BibitemOpen
  \href@noop {} {}\bibinfo {note} {See the Supplemental Material for
  theoretical and experimental details, as well as proofs of the main results,
  which includes references
  \cite{Lindblad,horowitz2,leggio,campisi,manzanoPRE,gong,liu,elouard,karimi,TR1,sagawaREV,Doob}.}\BibitemShut
  {Stop}%
\bibitem [{app()}]{apparent}%
  \BibitemOpen
  \href@noop {} {}\bibinfo {note} {Here we refer to apparent "second-law
  violations" since the second law can be restored by considering the erasure
  of the information acquired to implement the gambling strategy, as in other
  versions of Maxwell's demon.}\BibitemShut {Stop}%
\bibitem [{\citenamefont {Maillet}\ \emph {et~al.}(2019)\citenamefont
  {Maillet}, \citenamefont {Erdman}, \citenamefont {Cavina}, \citenamefont
  {Bhandari}, \citenamefont {Mannila}, \citenamefont {Peltonen}, \citenamefont
  {Mari}, \citenamefont {Taddei}, \citenamefont {Jarzynski}, \citenamefont
  {Giovannetti},\ and\ \citenamefont {Pekola}}]{expdata}%
  \BibitemOpen
  \bibfield  {author} {\bibinfo {author} {\bibfnamefont {O.}~\bibnamefont
  {Maillet}}, \bibinfo {author} {\bibfnamefont {P.~A.}\ \bibnamefont {Erdman}},
  \bibinfo {author} {\bibfnamefont {V.}~\bibnamefont {Cavina}}, \bibinfo
  {author} {\bibfnamefont {B.}~\bibnamefont {Bhandari}}, \bibinfo {author}
  {\bibfnamefont {E.~T.}\ \bibnamefont {Mannila}}, \bibinfo {author}
  {\bibfnamefont {J.~T.}\ \bibnamefont {Peltonen}}, \bibinfo {author}
  {\bibfnamefont {A.}~\bibnamefont {Mari}}, \bibinfo {author} {\bibfnamefont
  {F.}~\bibnamefont {Taddei}}, \bibinfo {author} {\bibfnamefont
  {C.}~\bibnamefont {Jarzynski}}, \bibinfo {author} {\bibfnamefont
  {V.}~\bibnamefont {Giovannetti}},\ and\ \bibinfo {author} {\bibfnamefont
  {J.}~\bibnamefont {Pekola}},\ }\bibfield  {title} {\emph {\bibinfo {title}
  {Optimal probabilistic work extraction beyond the free energy difference with
  a single-electron device}},\ }\href@noop {} {\bibfield  {journal} {\bibinfo
  {journal} {Phys. Rev. Lett.}\ }\textbf {\bibinfo {volume} {122}},\ \bibinfo
  {pages} {150604} (\bibinfo {year} {2019})}\BibitemShut {NoStop}%
\bibitem [{\citenamefont {Pekola}\ and\ \citenamefont
  {Khaymovich}(2019)}]{SETs}%
  \BibitemOpen
  \bibfield  {author} {\bibinfo {author} {\bibfnamefont {J.~P.}\ \bibnamefont
  {Pekola}}\ and\ \bibinfo {author} {\bibfnamefont {I.~M.}\ \bibnamefont
  {Khaymovich}},\ }\bibfield  {title} {\emph {\bibinfo {title} {Thermodynamics
  in single-electron circuits and superconducting qubits}},\ }\href@noop {}
  {\bibfield  {journal} {\bibinfo  {journal} {Annu. Rev. Condens. Mat. Phys.}\
  }\textbf {\bibinfo {volume} {10}},\ \bibinfo {pages} {193--212} (\bibinfo
  {year} {2019})}\BibitemShut {NoStop}%
\bibitem [{\citenamefont {Wiseman}\ and\ \citenamefont
  {Milburn}(2009)}]{milburn}%
  \BibitemOpen
  \bibfield  {author} {\bibinfo {author} {\bibfnamefont {H.~M.}\ \bibnamefont
  {Wiseman}}\ and\ \bibinfo {author} {\bibfnamefont {G.~J.}\ \bibnamefont
  {Milburn}},\ }\href@noop {} {\emph {\bibinfo {title} {Quantum measurement and
  control}}}\ (\bibinfo  {publisher} {Cambridge university press},\ \bibinfo
  {year} {2009})\BibitemShut {NoStop}%
\bibitem [{\citenamefont {Hekking}\ and\ \citenamefont
  {Pekola}(2013)}]{hekking}%
  \BibitemOpen
  \bibfield  {author} {\bibinfo {author} {\bibfnamefont {F.~W.~J.}\
  \bibnamefont {Hekking}}\ and\ \bibinfo {author} {\bibfnamefont {J.~P.}\
  \bibnamefont {Pekola}},\ }\bibfield  {title} {\emph {\bibinfo {title}
  {Quantum jump approach for work and dissipation in a two-level system}},\
  }\href@noop {} {\bibfield  {journal} {\bibinfo  {journal} {Phys. Rev. Lett.}\
  }\textbf {\bibinfo {volume} {111}},\ \bibinfo {pages} {093602} (\bibinfo
  {year} {2013})}\BibitemShut {NoStop}%
\bibitem [{\citenamefont {Horowitz}(2012)}]{horowitz}%
  \BibitemOpen
  \bibfield  {author} {\bibinfo {author} {\bibfnamefont {J.~M.}\ \bibnamefont
  {Horowitz}},\ }\bibfield  {title} {\emph {\bibinfo {title}
  {Quantum-trajectory approach to the stochastic thermodynamics of a forced
  harmonic oscillator}},\ }\href@noop {} {\bibfield  {journal} {\bibinfo
  {journal} {Phys. Rev. E}\ }\textbf {\bibinfo {volume} {85}},\ \bibinfo
  {pages} {031110} (\bibinfo {year} {2012})}\BibitemShut {NoStop}%
\bibitem [{\citenamefont {Manzano}\ \emph {et~al.}(2018)\citenamefont
  {Manzano}, \citenamefont {Horowitz},\ and\ \citenamefont
  {Parrondo}}]{manzano}%
  \BibitemOpen
  \bibfield  {author} {\bibinfo {author} {\bibfnamefont {G.}~\bibnamefont
  {Manzano}}, \bibinfo {author} {\bibfnamefont {J.~M.}\ \bibnamefont
  {Horowitz}},\ and\ \bibinfo {author} {\bibfnamefont {J.~M.~R.}\ \bibnamefont
  {Parrondo}},\ }\bibfield  {title} {\emph {\bibinfo {title} {Quantum
  fluctuation theorems for arbitrary environments: adiabatic and nonadiabatic
  entropy production}},\ }\href@noop {} {\bibfield  {journal} {\bibinfo
  {journal} {Phys. Rev. X}\ }\textbf {\bibinfo {volume} {8}},\ \bibinfo {pages}
  {031037} (\bibinfo {year} {2018})}\BibitemShut {NoStop}%
\bibitem [{\citenamefont {Mart{\'\i}nez}\ \emph {et~al.}(2016)\citenamefont
  {Mart{\'\i}nez}, \citenamefont {Rold{\'a}n}, \citenamefont {Dinis},
  \citenamefont {Petrov}, \citenamefont {Parrondo},\ and\ \citenamefont
  {Rica}}]{martinez2016brownian}%
  \BibitemOpen
  \bibfield  {author} {\bibinfo {author} {\bibfnamefont {I.~A.}\ \bibnamefont
  {Mart{\'\i}nez}}, \bibinfo {author} {\bibfnamefont {{\'E}.}~\bibnamefont
  {Rold{\'a}n}}, \bibinfo {author} {\bibfnamefont {L.}~\bibnamefont {Dinis}},
  \bibinfo {author} {\bibfnamefont {D.}~\bibnamefont {Petrov}}, \bibinfo
  {author} {\bibfnamefont {J.~M.}\ \bibnamefont {Parrondo}},\ and\ \bibinfo
  {author} {\bibfnamefont {R.~A.}\ \bibnamefont {Rica}},\ }\bibfield  {title}
  {\emph {\bibinfo {title} {Brownian carnot engine}},\ }\href@noop {}
  {\bibfield  {journal} {\bibinfo  {journal} {Nature Phys.}\ }\textbf {\bibinfo
  {volume} {12}},\ \bibinfo {pages} {67--70} (\bibinfo {year}
  {2016})}\BibitemShut {NoStop}%
\bibitem [{\citenamefont {Horowitz}\ and\ \citenamefont
  {Parrondo}(2011)}]{horowitz2011thermodynamic}%
  \BibitemOpen
  \bibfield  {author} {\bibinfo {author} {\bibfnamefont {J.~M.}\ \bibnamefont
  {Horowitz}}\ and\ \bibinfo {author} {\bibfnamefont {J.~M.~R.}\ \bibnamefont
  {Parrondo}},\ }\bibfield  {title} {\emph {\bibinfo {title} {Thermodynamic
  reversibility in feedback processes}},\ }\href@noop {} {\bibfield  {journal}
  {\bibinfo  {journal} {EPL}\ }\textbf {\bibinfo {volume} {95}},\ \bibinfo
  {pages} {10005} (\bibinfo {year} {2011})}\BibitemShut {NoStop}%
\bibitem [{\citenamefont {Falasco}\ and\ \citenamefont
  {Esposito}(2020)}]{PhysRevLett.125.120604}%
  \BibitemOpen
  \bibfield  {author} {\bibinfo {author} {\bibfnamefont {G.}~\bibnamefont
  {Falasco}}\ and\ \bibinfo {author} {\bibfnamefont {M.}~\bibnamefont
  {Esposito}},\ }\bibfield  {title} {\emph {\bibinfo {title} {Dissipation-time
  uncertainty relation}},\ }\href@noop {} {\bibfield  {journal} {\bibinfo
  {journal} {Phys. Rev. Lett.}\ }\textbf {\bibinfo {volume} {125}},\ \bibinfo
  {pages} {120604} (\bibinfo {year} {2020})}\BibitemShut {NoStop}%
\bibitem [{\citenamefont {Shiraishi}\ \emph {et~al.}(2018)\citenamefont
  {Shiraishi}, \citenamefont {Funo},\ and\ \citenamefont
  {Saito}}]{shiraishi2018speed}%
  \BibitemOpen
  \bibfield  {author} {\bibinfo {author} {\bibfnamefont {N.}~\bibnamefont
  {Shiraishi}}, \bibinfo {author} {\bibfnamefont {K.}~\bibnamefont {Funo}},\
  and\ \bibinfo {author} {\bibfnamefont {K.}~\bibnamefont {Saito}},\ }\bibfield
   {title} {\emph {\bibinfo {title} {Speed limit for classical stochastic
  processes}},\ }\href@noop {} {\bibfield  {journal} {\bibinfo  {journal}
  {Phys. Rev. Lett.}\ }\textbf {\bibinfo {volume} {121}},\ \bibinfo {pages}
  {070601} (\bibinfo {year} {2018})}\BibitemShut {NoStop}%
\bibitem [{\citenamefont {Minev}\ \emph {et~al.}(2019)\citenamefont {Minev},
  \citenamefont {Mundhada}, \citenamefont {Shankar}, \citenamefont {Reinhold},
  \citenamefont {Guti{\'e}rrez-J{\'a}uregui}, \citenamefont {Schoelkopf},
  \citenamefont {Mirrahimi}, \citenamefont {Carmichael},\ and\ \citenamefont
  {Devoret}}]{minev2019catch}%
  \BibitemOpen
  \bibfield  {author} {\bibinfo {author} {\bibfnamefont {Z.~K.}\ \bibnamefont
  {Minev}}, \bibinfo {author} {\bibfnamefont {S.~O.}\ \bibnamefont {Mundhada}},
  \bibinfo {author} {\bibfnamefont {S.}~\bibnamefont {Shankar}}, \bibinfo
  {author} {\bibfnamefont {P.}~\bibnamefont {Reinhold}}, \bibinfo {author}
  {\bibfnamefont {R.}~\bibnamefont {Guti{\'e}rrez-J{\'a}uregui}}, \bibinfo
  {author} {\bibfnamefont {R.~J.}\ \bibnamefont {Schoelkopf}}, \bibinfo
  {author} {\bibfnamefont {M.}~\bibnamefont {Mirrahimi}}, \bibinfo {author}
  {\bibfnamefont {H.~J.}\ \bibnamefont {Carmichael}},\ and\ \bibinfo {author}
  {\bibfnamefont {M.~H.}\ \bibnamefont {Devoret}},\ }\bibfield  {title} {\emph
  {\bibinfo {title} {To catch and reverse a quantum jump mid-flight}},\
  }\href@noop {} {\bibfield  {journal} {\bibinfo  {journal} {Nature}\ }\textbf
  {\bibinfo {volume} {570}},\ \bibinfo {pages} {200--204} (\bibinfo {year}
  {2019})}\BibitemShut {NoStop}%
\bibitem [{\citenamefont {Murch}\ \emph {et~al.}(2013)\citenamefont {Murch},
  \citenamefont {Weber}, \citenamefont {Macklin},\ and\ \citenamefont
  {Siddiqi}}]{Murch}%
  \BibitemOpen
  \bibfield  {author} {\bibinfo {author} {\bibfnamefont {K.~W.}\ \bibnamefont
  {Murch}}, \bibinfo {author} {\bibfnamefont {S.}~\bibnamefont {Weber}},
  \bibinfo {author} {\bibfnamefont {C.}~\bibnamefont {Macklin}},\ and\ \bibinfo
  {author} {\bibfnamefont {I.}~\bibnamefont {Siddiqi}},\ }\bibfield  {title}
  {\emph {\bibinfo {title} {Observing single quantum trajectories of a
  superconducting quantum}},\ }\href@noop {} {\bibfield  {journal} {\bibinfo
  {journal} {Nature}\ }\textbf {\bibinfo {volume} {502}},\ \bibinfo {pages}
  {211--214} (\bibinfo {year} {2013})}\BibitemShut {NoStop}%
\bibitem [{\citenamefont {Dinis}\ \emph {et~al.}(2020)\citenamefont {Dinis},
  \citenamefont {Unterberger},\ and\ \citenamefont {Lacoste}}]{Dinis}%
  \BibitemOpen
  \bibfield  {author} {\bibinfo {author} {\bibfnamefont {L.}~\bibnamefont
  {Dinis}}, \bibinfo {author} {\bibfnamefont {J.}~\bibnamefont {Unterberger}},\
  and\ \bibinfo {author} {\bibfnamefont {D.}~\bibnamefont {Lacoste}},\
  }\bibfield  {title} {\emph {\bibinfo {title} {Phase transitions in optimal
  strategies for betting}},\ }\href@noop {} {\bibfield  {journal} {\bibinfo
  {journal} {arXiv:2005.11698}\ } (\bibinfo {year} {2020})}\BibitemShut
  {NoStop}%
\bibitem [{\citenamefont {Ito}(2016)}]{Ito}%
  \BibitemOpen
  \bibfield  {author} {\bibinfo {author} {\bibfnamefont {S.}~\bibnamefont
  {Ito}},\ }\bibfield  {title} {\emph {\bibinfo {title} {Backward transfer
  entropy: Informational measure for detecting hidden markov models and its
  interpretations in thermodynamics, gambling and causality}},\ }\href@noop {}
  {\bibfield  {journal} {\bibinfo  {journal} {Sci. Rep.}\ }\textbf {\bibinfo
  {volume} {6}},\ \bibinfo {pages} {36831} (\bibinfo {year}
  {2016})}\BibitemShut {NoStop}%
\bibitem [{\citenamefont {Harmer}\ and\ \citenamefont {Abbott}(1999)}]{Harmer}%
  \BibitemOpen
  \bibfield  {author} {\bibinfo {author} {\bibfnamefont {G.~P.}\ \bibnamefont
  {Harmer}}\ and\ \bibinfo {author} {\bibfnamefont {D.}~\bibnamefont
  {Abbott}},\ }\bibfield  {title} {\emph {\bibinfo {title} {Losing strategies
  can win by {P}arrondo's paradox}},\ }\href@noop {} {\bibfield  {journal}
  {\bibinfo  {journal} {Nature}\ }\textbf {\bibinfo {volume} {402}},\ \bibinfo
  {pages} {864--864} (\bibinfo {year} {1999})}\BibitemShut {NoStop}%
\bibitem [{\citenamefont {Lindblad}(1976)}]{Lindblad}%
  \BibitemOpen
  \bibfield  {author} {\bibinfo {author} {\bibfnamefont {G.}~\bibnamefont
  {Lindblad}},\ }\bibfield  {title} {\emph {\bibinfo {title} {On the generators
  of quantum dynamical semigroups}},\ }\href@noop {} {\bibfield  {journal}
  {\bibinfo  {journal} {Comms. Math. Phys.}\ }\textbf {\bibinfo {volume}
  {48}},\ \bibinfo {pages} {119--130} (\bibinfo {year} {1976})}\BibitemShut
  {NoStop}%
\bibitem [{\citenamefont {Horowitz}\ and\ \citenamefont
  {Parrondo}(2013)}]{horowitz2}%
  \BibitemOpen
  \bibfield  {author} {\bibinfo {author} {\bibfnamefont {J.~M.}\ \bibnamefont
  {Horowitz}}\ and\ \bibinfo {author} {\bibfnamefont {J.~M.~R.}\ \bibnamefont
  {Parrondo}},\ }\bibfield  {title} {\emph {\bibinfo {title} {Entropy
  production along nonequilibrium quantum jump trajectories}},\ }\href@noop {}
  {\bibfield  {journal} {\bibinfo  {journal} {New J. Phys.}\ }\textbf {\bibinfo
  {volume} {15}},\ \bibinfo {pages} {085028} (\bibinfo {year}
  {2013})}\BibitemShut {NoStop}%
\bibitem [{\citenamefont {Leggio}\ \emph {et~al.}(2013)\citenamefont {Leggio},
  \citenamefont {Napoli}, \citenamefont {Messina},\ and\ \citenamefont
  {Breuer}}]{leggio}%
  \BibitemOpen
  \bibfield  {author} {\bibinfo {author} {\bibfnamefont {B.}~\bibnamefont
  {Leggio}}, \bibinfo {author} {\bibfnamefont {A.}~\bibnamefont {Napoli}},
  \bibinfo {author} {\bibfnamefont {A.}~\bibnamefont {Messina}},\ and\ \bibinfo
  {author} {\bibfnamefont {H.-P.}\ \bibnamefont {Breuer}},\ }\bibfield  {title}
  {\emph {\bibinfo {title} {Entropy production and information fluctuations
  along quantum trajectories}},\ }\href@noop {} {\bibfield  {journal} {\bibinfo
   {journal} {Phys. Rev. A}\ }\textbf {\bibinfo {volume} {88}},\ \bibinfo
  {pages} {042111} (\bibinfo {year} {2013})}\BibitemShut {NoStop}%
\bibitem [{\citenamefont {Campisi}\ \emph {et~al.}(2015)\citenamefont
  {Campisi}, \citenamefont {Pekola},\ and\ \citenamefont {Fazio}}]{campisi}%
  \BibitemOpen
  \bibfield  {author} {\bibinfo {author} {\bibfnamefont {M.}~\bibnamefont
  {Campisi}}, \bibinfo {author} {\bibfnamefont {J.~P.}\ \bibnamefont
  {Pekola}},\ and\ \bibinfo {author} {\bibfnamefont {R.}~\bibnamefont
  {Fazio}},\ }\bibfield  {title} {\emph {\bibinfo {title} {Nonequilibrium
  fluctuations in quantum heat engines: theory, example, and possible solid
  state experiments}},\ }\href@noop {} {\bibfield  {journal} {\bibinfo
  {journal} {New J. Phys.}\ }\textbf {\bibinfo {volume} {17}},\ \bibinfo
  {pages} {035012} (\bibinfo {year} {2015})}\BibitemShut {NoStop}%
\bibitem [{\citenamefont {Manzano}\ \emph {et~al.}(2015)\citenamefont
  {Manzano}, \citenamefont {Horowitz},\ and\ \citenamefont
  {Parrondo}}]{manzanoPRE}%
  \BibitemOpen
  \bibfield  {author} {\bibinfo {author} {\bibfnamefont {G.}~\bibnamefont
  {Manzano}}, \bibinfo {author} {\bibfnamefont {J.~M.}\ \bibnamefont
  {Horowitz}},\ and\ \bibinfo {author} {\bibfnamefont {J.~M.~R.}\ \bibnamefont
  {Parrondo}},\ }\bibfield  {title} {\emph {\bibinfo {title} {Nonequilibrium
  potential and fluctuation theorems for quantum maps}},\ }\href@noop {}
  {\bibfield  {journal} {\bibinfo  {journal} {Phys. Rev. E}\ }\textbf {\bibinfo
  {volume} {92}},\ \bibinfo {pages} {032129} (\bibinfo {year}
  {2015})}\BibitemShut {NoStop}%
\bibitem [{\citenamefont {Gong}\ \emph {et~al.}(2016)\citenamefont {Gong},
  \citenamefont {Ashida},\ and\ \citenamefont {Ueda}}]{gong}%
  \BibitemOpen
  \bibfield  {author} {\bibinfo {author} {\bibfnamefont {Z.}~\bibnamefont
  {Gong}}, \bibinfo {author} {\bibfnamefont {Y.}~\bibnamefont {Ashida}},\ and\
  \bibinfo {author} {\bibfnamefont {M.}~\bibnamefont {Ueda}},\ }\bibfield
  {title} {\emph {\bibinfo {title} {Quantum-trajectory thermodynamics with
  discrete feedback control}},\ }\href@noop {} {\bibfield  {journal} {\bibinfo
  {journal} {Phys. Rev. A}\ }\textbf {\bibinfo {volume} {94}},\ \bibinfo
  {pages} {012107} (\bibinfo {year} {2016})}\BibitemShut {NoStop}%
\bibitem [{\citenamefont {Liu}\ and\ \citenamefont {Xi}(2016)}]{liu}%
  \BibitemOpen
  \bibfield  {author} {\bibinfo {author} {\bibfnamefont {F.}~\bibnamefont
  {Liu}}\ and\ \bibinfo {author} {\bibfnamefont {J.}~\bibnamefont {Xi}},\
  }\bibfield  {title} {\emph {\bibinfo {title} {Characteristic functions based
  on a quantum jump trajectory}},\ }\href@noop {} {\bibfield  {journal}
  {\bibinfo  {journal} {Phys. Rev. E}\ }\textbf {\bibinfo {volume} {94}},\
  \bibinfo {pages} {062133} (\bibinfo {year} {2016})}\BibitemShut {NoStop}%
\bibitem [{\citenamefont {Elouard}\ \emph {et~al.}(2017)\citenamefont
  {Elouard}, \citenamefont {Herrera-Mart\'{\i}}, \citenamefont {Clusel},\ and\
  \citenamefont {Auff{\`e}ves}}]{elouard}%
  \BibitemOpen
  \bibfield  {author} {\bibinfo {author} {\bibfnamefont {C.}~\bibnamefont
  {Elouard}}, \bibinfo {author} {\bibfnamefont {D.~A.}\ \bibnamefont
  {Herrera-Mart\'{\i}}}, \bibinfo {author} {\bibfnamefont {M.}~\bibnamefont
  {Clusel}},\ and\ \bibinfo {author} {\bibfnamefont {A.}~\bibnamefont
  {Auff{\`e}ves}},\ }\bibfield  {title} {\emph {\bibinfo {title} {The role of
  quantum measurement in stochastic thermodynamics}},\ }\href@noop {}
  {\bibfield  {journal} {\bibinfo  {journal} {npj Quant. Info.}\ }\textbf
  {\bibinfo {volume} {3}},\ \bibinfo {pages} {9} (\bibinfo {year}
  {2017})}\BibitemShut {NoStop}%
\bibitem [{\citenamefont {Karimi}\ and\ \citenamefont {Pekola}(2020)}]{karimi}%
  \BibitemOpen
  \bibfield  {author} {\bibinfo {author} {\bibfnamefont {B.}~\bibnamefont
  {Karimi}}\ and\ \bibinfo {author} {\bibfnamefont {J.~P.}\ \bibnamefont
  {Pekola}},\ }\bibfield  {title} {\emph {\bibinfo {title} {Quantum trajectory
  analysis of single microwave photon detection by nanocalorimetry}},\
  }\href@noop {} {\bibfield  {journal} {\bibinfo  {journal} {Phys. Rev. Lett.}\
  }\textbf {\bibinfo {volume} {124}},\ \bibinfo {pages} {170601} (\bibinfo
  {year} {2020})}\BibitemShut {NoStop}%
\bibitem [{\citenamefont {Kawai}\ \emph {et~al.}(2007)\citenamefont {Kawai},
  \citenamefont {Parrondo},\ and\ \citenamefont {{Van den Broeck}}}]{TR1}%
  \BibitemOpen
  \bibfield  {author} {\bibinfo {author} {\bibfnamefont {R.}~\bibnamefont
  {Kawai}}, \bibinfo {author} {\bibfnamefont {J.~M.~R.}\ \bibnamefont
  {Parrondo}},\ and\ \bibinfo {author} {\bibfnamefont {C.}~\bibnamefont {{Van
  den Broeck}}},\ }\bibfield  {title} {\emph {\bibinfo {title} {Dissipation:
  The phase-space perspective}},\ }\href@noop {} {\bibfield  {journal}
  {\bibinfo  {journal} {Phys. Rev. Lett.}\ }\textbf {\bibinfo {volume} {98}},\
  \bibinfo {pages} {080602} (\bibinfo {year} {2007})}\BibitemShut {NoStop}%
\bibitem [{\citenamefont {Sagawa}(2012)}]{sagawaREV}%
  \BibitemOpen
  \bibfield  {author} {\bibinfo {author} {\bibfnamefont {T.}~\bibnamefont
  {Sagawa}},\ }\href@noop {} {\emph {\bibinfo {title} {Second Law-Like
  Inequalities with Quantum Relative Entropy: An Introduction in {L}ectures on
  Quantum Computing, Thermodynamics and Statistical Physics}}},\ edited by\
  \bibinfo {editor} {\bibfnamefont {M.}~\bibnamefont {Nakahara}}\ and\ \bibinfo
  {editor} {\bibfnamefont {S.}~\bibnamefont {Tanaka}},\ Kinki University Series
  on Quantum Computing\ (\bibinfo  {publisher} {World Scientific},\ \bibinfo
  {year} {2012})\BibitemShut {NoStop}%
\bibitem [{\citenamefont {Doob}(1953)}]{Doob}%
  \BibitemOpen
  \bibfield  {author} {\bibinfo {author} {\bibfnamefont {J.}~\bibnamefont
  {Doob}},\ }\href@noop {} {\emph {\bibinfo {title} {Stochastic Processes}}}\
  (\bibinfo  {publisher} {John Wiley and Sons},\ \bibinfo {year}
  {1953})\BibitemShut {NoStop}%
\end{thebibliography}

\begin{thebibliography}{24}%
\makeatletter
\providecommand \@ifxundefined [1]{%
 \@ifx{#1\undefined}
}%
\providecommand \@ifnum [1]{%
 \ifnum #1\expandafter \@firstoftwo
 \else \expandafter \@secondoftwo
 \fi
}%
\providecommand \@ifx [1]{%
 \ifx #1\expandafter \@firstoftwo
 \else \expandafter \@secondoftwo
 \fi
}%
\providecommand \natexlab [1]{#1}%
\providecommand \enquote  [1]{``#1''}%
\providecommand \bibnamefont  [1]{#1}%
\providecommand \bibfnamefont [1]{#1}%
\providecommand \citenamefont [1]{#1}%
\providecommand \href@noop [0]{\@secondoftwo}%
\providecommand \href [0]{\begingroup \@sanitize@url \@href}%
\providecommand \@href[1]{\@@startlink{#1}\@@href}%
\providecommand \@@href[1]{\endgroup#1\@@endlink}%
\providecommand \@sanitize@url [0]{\catcode `\\12\catcode `\$12\catcode
  `\&12\catcode `\#12\catcode `\^12\catcode `\_12\catcode `\%12\relax}%
\providecommand \@@startlink[1]{}%
\providecommand \@@endlink[0]{}%
\providecommand \url  [0]{\begingroup\@sanitize@url \@url }%
\providecommand \@url [1]{\endgroup\@href {#1}{\urlprefix }}%
\providecommand \urlprefix  [0]{URL }%
\providecommand \Eprint [0]{\href }%
\providecommand \doibase [0]{https://doi.org/}%
\providecommand \selectlanguage [0]{\@gobble}%
\providecommand \bibinfo  [0]{\@secondoftwo}%
\providecommand \bibfield  [0]{\@secondoftwo}%
\providecommand \translation [1]{[#1]}%
\providecommand \BibitemOpen [0]{}%
\providecommand \bibitemStop [0]{}%
\providecommand \bibitemNoStop [0]{.\EOS\space}%
\providecommand \EOS [0]{\spacefactor3000\relax}%
\providecommand \BibitemShut  [1]{\csname bibitem#1\endcsname}%
\let\auto@bib@innerbib\@empty
\bibitem [{\citenamefont {Maillet}\ \emph {et~al.}(2019)\citenamefont
  {Maillet}, \citenamefont {Erdman}, \citenamefont {Cavina}, \citenamefont
  {Bhandari}, \citenamefont {Mannila}, \citenamefont {Peltonen}, \citenamefont
  {Mari}, \citenamefont {Taddei}, \citenamefont {Jarzynski}, \citenamefont
  {Giovannetti},\ and\ \citenamefont {Pekola}}]{Sexpdata}%
  \BibitemOpen
  \bibfield  {author} {\bibinfo {author} {\bibfnamefont {O.}~\bibnamefont
  {Maillet}}, \bibinfo {author} {\bibfnamefont {P.~A.}\ \bibnamefont {Erdman}},
  \bibinfo {author} {\bibfnamefont {V.}~\bibnamefont {Cavina}}, \bibinfo
  {author} {\bibfnamefont {B.}~\bibnamefont {Bhandari}}, \bibinfo {author}
  {\bibfnamefont {E.~T.}\ \bibnamefont {Mannila}}, \bibinfo {author}
  {\bibfnamefont {J.~T.}\ \bibnamefont {Peltonen}}, \bibinfo {author}
  {\bibfnamefont {A.}~\bibnamefont {Mari}}, \bibinfo {author} {\bibfnamefont
  {F.}~\bibnamefont {Taddei}}, \bibinfo {author} {\bibfnamefont
  {C.}~\bibnamefont {Jarzynski}}, \bibinfo {author} {\bibfnamefont
  {V.}~\bibnamefont {Giovannetti}},\ and\ \bibinfo {author} {\bibfnamefont
  {J.}~\bibnamefont {Pekola}},\ }\bibfield  {title} {\emph {\bibinfo {title}
  {Optimal probabilistic work extraction beyond the free energy difference with
  a single-electron device}},\ }\href@noop {} {\bibfield  {journal} {\bibinfo
  {journal} {Phys. Rev. Lett.}\ }\textbf {\bibinfo {volume} {122}},\ \bibinfo
  {pages} {150604} (\bibinfo {year} {2019})}\BibitemShut {NoStop}%
\bibitem [{\citenamefont {Averin}\ \emph {et~al.}(1991)\citenamefont {Averin},
  \citenamefont {Korotkov},\ and\ \citenamefont {Likharev}}]{PhysRevB.44.6199}%
  \BibitemOpen
  \bibfield  {author} {\bibinfo {author} {\bibfnamefont {D.~V.}\ \bibnamefont
  {Averin}}, \bibinfo {author} {\bibfnamefont {A.~N.}\ \bibnamefont
  {Korotkov}},\ and\ \bibinfo {author} {\bibfnamefont {K.~K.}\ \bibnamefont
  {Likharev}},\ }\bibfield  {title} {\emph {\bibinfo {title} {Theory of
  single-electron charging of quantum wells and dots}},\ }\href@noop {}
  {\bibfield  {journal} {\bibinfo  {journal} {Phys. Rev. B}\ }\textbf {\bibinfo
  {volume} {44}},\ \bibinfo {pages} {6199--6211} (\bibinfo {year}
  {1991})}\BibitemShut {NoStop}%
\bibitem [{\citenamefont {Wiseman}\ and\ \citenamefont
  {Milburn}(2009)}]{Smilburn}%
  \BibitemOpen
  \bibfield  {author} {\bibinfo {author} {\bibfnamefont {H.~M.}\ \bibnamefont
  {Wiseman}}\ and\ \bibinfo {author} {\bibfnamefont {G.~J.}\ \bibnamefont
  {Milburn}},\ }\href@noop {} {\emph {\bibinfo {title} {Quantum measurement and
  control}}}\ (\bibinfo  {publisher} {Cambridge university press},\ \bibinfo
  {year} {2009})\BibitemShut {NoStop}%
\bibitem [{\citenamefont {Lindblad}(1976)}]{SLindblad}%
  \BibitemOpen
  \bibfield  {author} {\bibinfo {author} {\bibfnamefont {G.}~\bibnamefont
  {Lindblad}},\ }\bibfield  {title} {\emph {\bibinfo {title} {On the generators
  of quantum dynamical semigroups}},\ }\href@noop {} {\bibfield  {journal}
  {\bibinfo  {journal} {Comms. Math. Phys.}\ }\textbf {\bibinfo {volume}
  {48}},\ \bibinfo {pages} {119--130} (\bibinfo {year} {1976})}\BibitemShut
  {NoStop}%
\bibitem [{\citenamefont {Horowitz}(2012)}]{Shorowitz}%
  \BibitemOpen
  \bibfield  {author} {\bibinfo {author} {\bibfnamefont {J.~M.}\ \bibnamefont
  {Horowitz}},\ }\bibfield  {title} {\emph {\bibinfo {title}
  {Quantum-trajectory approach to the stochastic thermodynamics of a forced
  harmonic oscillator}},\ }\href@noop {} {\bibfield  {journal} {\bibinfo
  {journal} {Phys. Rev. E}\ }\textbf {\bibinfo {volume} {85}},\ \bibinfo
  {pages} {031110} (\bibinfo {year} {2012})}\BibitemShut {NoStop}%
\bibitem [{\citenamefont {Hekking}\ and\ \citenamefont
  {Pekola}(2013)}]{Shekking}%
  \BibitemOpen
  \bibfield  {author} {\bibinfo {author} {\bibfnamefont {F.~W.~J.}\
  \bibnamefont {Hekking}}\ and\ \bibinfo {author} {\bibfnamefont {J.~P.}\
  \bibnamefont {Pekola}},\ }\bibfield  {title} {\emph {\bibinfo {title}
  {Quantum jump approach for work and dissipation in a two-level system}},\
  }\href@noop {} {\bibfield  {journal} {\bibinfo  {journal} {Phys. Rev. Lett.}\
  }\textbf {\bibinfo {volume} {111}},\ \bibinfo {pages} {093602} (\bibinfo
  {year} {2013})}\BibitemShut {NoStop}%
\bibitem [{\citenamefont {Horowitz}\ and\ \citenamefont
  {Parrondo}(2013)}]{Shorowitz2}%
  \BibitemOpen
  \bibfield  {author} {\bibinfo {author} {\bibfnamefont {J.~M.}\ \bibnamefont
  {Horowitz}}\ and\ \bibinfo {author} {\bibfnamefont {J.~M.~R.}\ \bibnamefont
  {Parrondo}},\ }\bibfield  {title} {\emph {\bibinfo {title} {Entropy
  production along nonequilibrium quantum jump trajectories}},\ }\href@noop {}
  {\bibfield  {journal} {\bibinfo  {journal} {New J. Phys.}\ }\textbf {\bibinfo
  {volume} {15}},\ \bibinfo {pages} {085028} (\bibinfo {year}
  {2013})}\BibitemShut {NoStop}%
\bibitem [{\citenamefont {Leggio}\ \emph {et~al.}(2013)\citenamefont {Leggio},
  \citenamefont {Napoli}, \citenamefont {Messina},\ and\ \citenamefont
  {Breuer}}]{Sleggio}%
  \BibitemOpen
  \bibfield  {author} {\bibinfo {author} {\bibfnamefont {B.}~\bibnamefont
  {Leggio}}, \bibinfo {author} {\bibfnamefont {A.}~\bibnamefont {Napoli}},
  \bibinfo {author} {\bibfnamefont {A.}~\bibnamefont {Messina}},\ and\ \bibinfo
  {author} {\bibfnamefont {H.-P.}\ \bibnamefont {Breuer}},\ }\bibfield  {title}
  {\emph {\bibinfo {title} {Entropy production and information fluctuations
  along quantum trajectories}},\ }\href@noop {} {\bibfield  {journal} {\bibinfo
   {journal} {Phys. Rev. A}\ }\textbf {\bibinfo {volume} {88}},\ \bibinfo
  {pages} {042111} (\bibinfo {year} {2013})}\BibitemShut {NoStop}%
\bibitem [{\citenamefont {Campisi}\ \emph {et~al.}(2015)\citenamefont
  {Campisi}, \citenamefont {Pekola},\ and\ \citenamefont {Fazio}}]{Scampisi}%
  \BibitemOpen
  \bibfield  {author} {\bibinfo {author} {\bibfnamefont {M.}~\bibnamefont
  {Campisi}}, \bibinfo {author} {\bibfnamefont {J.~P.}\ \bibnamefont
  {Pekola}},\ and\ \bibinfo {author} {\bibfnamefont {R.}~\bibnamefont
  {Fazio}},\ }\bibfield  {title} {\emph {\bibinfo {title} {Nonequilibrium
  fluctuations in quantum heat engines: theory, example, and possible solid
  state experiments}},\ }\href@noop {} {\bibfield  {journal} {\bibinfo
  {journal} {New J. Phys.}\ }\textbf {\bibinfo {volume} {17}},\ \bibinfo
  {pages} {035012} (\bibinfo {year} {2015})}\BibitemShut {NoStop}%
\bibitem [{\citenamefont {Manzano}\ \emph {et~al.}(2015)\citenamefont
  {Manzano}, \citenamefont {Horowitz},\ and\ \citenamefont
  {Parrondo}}]{SmanzanoPRE}%
  \BibitemOpen
  \bibfield  {author} {\bibinfo {author} {\bibfnamefont {G.}~\bibnamefont
  {Manzano}}, \bibinfo {author} {\bibfnamefont {J.~M.}\ \bibnamefont
  {Horowitz}},\ and\ \bibinfo {author} {\bibfnamefont {J.~M.~R.}\ \bibnamefont
  {Parrondo}},\ }\bibfield  {title} {\emph {\bibinfo {title} {Nonequilibrium
  potential and fluctuation theorems for quantum maps}},\ }\href@noop {}
  {\bibfield  {journal} {\bibinfo  {journal} {Phys. Rev. E}\ }\textbf {\bibinfo
  {volume} {92}},\ \bibinfo {pages} {032129} (\bibinfo {year}
  {2015})}\BibitemShut {NoStop}%
\bibitem [{\citenamefont {Gong}\ \emph {et~al.}(2016)\citenamefont {Gong},
  \citenamefont {Ashida},\ and\ \citenamefont {Ueda}}]{Sgong}%
  \BibitemOpen
  \bibfield  {author} {\bibinfo {author} {\bibfnamefont {Z.}~\bibnamefont
  {Gong}}, \bibinfo {author} {\bibfnamefont {Y.}~\bibnamefont {Ashida}},\ and\
  \bibinfo {author} {\bibfnamefont {M.}~\bibnamefont {Ueda}},\ }\bibfield
  {title} {\emph {\bibinfo {title} {Quantum-trajectory thermodynamics with
  discrete feedback control}},\ }\href@noop {} {\bibfield  {journal} {\bibinfo
  {journal} {Phys. Rev. A}\ }\textbf {\bibinfo {volume} {94}},\ \bibinfo
  {pages} {012107} (\bibinfo {year} {2016})}\BibitemShut {NoStop}%
\bibitem [{\citenamefont {Liu}\ and\ \citenamefont {Xi}(2016)}]{Sliu}%
  \BibitemOpen
  \bibfield  {author} {\bibinfo {author} {\bibfnamefont {F.}~\bibnamefont
  {Liu}}\ and\ \bibinfo {author} {\bibfnamefont {J.}~\bibnamefont {Xi}},\
  }\bibfield  {title} {\emph {\bibinfo {title} {Characteristic functions based
  on a quantum jump trajectory}},\ }\href@noop {} {\bibfield  {journal}
  {\bibinfo  {journal} {Phys. Rev. E}\ }\textbf {\bibinfo {volume} {94}},\
  \bibinfo {pages} {062133} (\bibinfo {year} {2016})}\BibitemShut {NoStop}%
\bibitem [{\citenamefont {Elouard}\ \emph {et~al.}(2017)\citenamefont
  {Elouard}, \citenamefont {Herrera-Mart\'{\i}}, \citenamefont {Clusel},\ and\
  \citenamefont {Auff{\`e}ves}}]{Selouard}%
  \BibitemOpen
  \bibfield  {author} {\bibinfo {author} {\bibfnamefont {C.}~\bibnamefont
  {Elouard}}, \bibinfo {author} {\bibfnamefont {D.~A.}\ \bibnamefont
  {Herrera-Mart\'{\i}}}, \bibinfo {author} {\bibfnamefont {M.}~\bibnamefont
  {Clusel}},\ and\ \bibinfo {author} {\bibfnamefont {A.}~\bibnamefont
  {Auff{\`e}ves}},\ }\bibfield  {title} {\emph {\bibinfo {title} {The role of
  quantum measurement in stochastic thermodynamics}},\ }\href@noop {}
  {\bibfield  {journal} {\bibinfo  {journal} {npj Quant. Info.}\ }\textbf
  {\bibinfo {volume} {3}},\ \bibinfo {pages} {9} (\bibinfo {year}
  {2017})}\BibitemShut {NoStop}%
\bibitem [{\citenamefont {Manzano}\ \emph {et~al.}(2018)\citenamefont
  {Manzano}, \citenamefont {Horowitz},\ and\ \citenamefont
  {Parrondo}}]{Smanzano}%
  \BibitemOpen
  \bibfield  {author} {\bibinfo {author} {\bibfnamefont {G.}~\bibnamefont
  {Manzano}}, \bibinfo {author} {\bibfnamefont {J.~M.}\ \bibnamefont
  {Horowitz}},\ and\ \bibinfo {author} {\bibfnamefont {J.~M.~R.}\ \bibnamefont
  {Parrondo}},\ }\bibfield  {title} {\emph {\bibinfo {title} {Quantum
  fluctuation theorems for arbitrary environments: adiabatic and nonadiabatic
  entropy production}},\ }\href@noop {} {\bibfield  {journal} {\bibinfo
  {journal} {Phys. Rev. X}\ }\textbf {\bibinfo {volume} {8}},\ \bibinfo {pages}
  {031037} (\bibinfo {year} {2018})}\BibitemShut {NoStop}%
\bibitem [{\citenamefont {Karimi}\ and\ \citenamefont {Pekola}(2020)}]{Skarimi}%
  \BibitemOpen
  \bibfield  {author} {\bibinfo {author} {\bibfnamefont {B.}~\bibnamefont
  {Karimi}}\ and\ \bibinfo {author} {\bibfnamefont {J.~P.}\ \bibnamefont
  {Pekola}},\ }\bibfield  {title} {\emph {\bibinfo {title} {Quantum trajectory
  analysis of single microwave photon detection by nanocalorimetry}},\
  }\href@noop {} {\bibfield  {journal} {\bibinfo  {journal} {Phys. Rev. Lett.}\
  }\textbf {\bibinfo {volume} {124}},\ \bibinfo {pages} {170601} (\bibinfo
  {year} {2020})}\BibitemShut {NoStop}%
\bibitem [{\citenamefont {Williams}(1991)}]{SWilliams}%
  \BibitemOpen
  \bibfield  {author} {\bibinfo {author} {\bibfnamefont {D.}~\bibnamefont
  {Williams}},\ }\href@noop {} {\emph {\bibinfo {title} {Probability with
  martingales}}}\ (\bibinfo  {publisher} {Cambridge university press},\
  \bibinfo {year} {1991})\BibitemShut {NoStop}%
\bibitem [{\citenamefont {Neri}\ \emph {et~al.}(2017)\citenamefont {Neri},
  \citenamefont {Rold{\'a}n},\ and\ \citenamefont {J{\"u}licher}}]{Sneri}%
  \BibitemOpen
  \bibfield  {author} {\bibinfo {author} {\bibfnamefont {I.}~\bibnamefont
  {Neri}}, \bibinfo {author} {\bibfnamefont {{\'E}.}~\bibnamefont
  {Rold{\'a}n}},\ and\ \bibinfo {author} {\bibfnamefont {F.}~\bibnamefont
  {J{\"u}licher}},\ }\bibfield  {title} {\emph {\bibinfo {title} {Statistics of
  infima and stopping times of entropy production and applications to active
  molecular processes}},\ }\href@noop {} {\bibfield  {journal} {\bibinfo
  {journal} {Phys. Rev. X}\ }\textbf {\bibinfo {volume} {7}},\ \bibinfo {pages}
  {011019} (\bibinfo {year} {2017})}\BibitemShut {NoStop}%
\bibitem [{\citenamefont {Chetrite}\ and\ \citenamefont
  {Gupta}(2011)}]{Sraphaelshamik}%
  \BibitemOpen
  \bibfield  {author} {\bibinfo {author} {\bibfnamefont {R.}~\bibnamefont
  {Chetrite}}\ and\ \bibinfo {author} {\bibfnamefont {S.}~\bibnamefont
  {Gupta}},\ }\bibfield  {title} {\emph {\bibinfo {title} {Two refreshing views
  of fluctuation theorems through kinematics elements and exponential
  martingale}},\ }\href@noop {} {\bibfield  {journal} {\bibinfo  {journal} {J.
  Stat. Phys.}\ }\textbf {\bibinfo {volume} {143}},\ \bibinfo {pages} {543}
  (\bibinfo {year} {2011})}\BibitemShut {NoStop}%
\bibitem [{\citenamefont {Kawai}\ \emph {et~al.}(2007)\citenamefont {Kawai},
  \citenamefont {Parrondo},\ and\ \citenamefont {{Van den Broeck}}}]{STR1}%
  \BibitemOpen
  \bibfield  {author} {\bibinfo {author} {\bibfnamefont {R.}~\bibnamefont
  {Kawai}}, \bibinfo {author} {\bibfnamefont {J.~M.~R.}\ \bibnamefont
  {Parrondo}},\ and\ \bibinfo {author} {\bibfnamefont {C.}~\bibnamefont {{Van
  den Broeck}}},\ }\bibfield  {title} {\emph {\bibinfo {title} {Dissipation:
  The phase-space perspective}},\ }\href@noop {} {\bibfield  {journal}
  {\bibinfo  {journal} {Phys. Rev. Lett.}\ }\textbf {\bibinfo {volume} {98}},\
  \bibinfo {pages} {080602} (\bibinfo {year} {2007})}\BibitemShut {NoStop}%
\bibitem [{\citenamefont {Sagawa}(2012)}]{SsagawaREV}%
  \BibitemOpen
  \bibfield  {author} {\bibinfo {author} {\bibfnamefont {T.}~\bibnamefont
  {Sagawa}},\ }\href@noop {} {\emph {\bibinfo {title} {Second Law-Like
  Inequalities with Quantum Relative Entropy: An Introduction in {L}ectures on
  Quantum Computing, Thermodynamics and Statistical Physics}}},\ edited by\
  \bibinfo {editor} {\bibfnamefont {M.}~\bibnamefont {Nakahara}}\ and\ \bibinfo
  {editor} {\bibfnamefont {S.}~\bibnamefont {Tanaka}},\ Kinki University Series
  on Quantum Computing\ (\bibinfo  {publisher} {World Scientific},\ \bibinfo
  {year} {2012})\BibitemShut {NoStop}%
\bibitem [{\citenamefont {Doob}(1953)}]{SDoob}%
  \BibitemOpen
  \bibfield  {author} {\bibinfo {author} {\bibfnamefont {J.}~\bibnamefont
  {Doob}},\ }\href@noop {} {\emph {\bibinfo {title} {Stochastic Processes}}}\
  (\bibinfo  {publisher} {John Wiley and Sons},\ \bibinfo {year}
  {1953})\BibitemShut {NoStop}%
\bibitem [{\citenamefont {Seifert}(2012)}]{SSeifertREV}%
  \BibitemOpen
  \bibfield  {author} {\bibinfo {author} {\bibfnamefont {U.}~\bibnamefont
  {Seifert}},\ }\bibfield  {title} {\emph {\bibinfo {title} {Stochastic
  thermodynamics, fluctuation theorems and molecular machines}},\ }\href@noop
  {} {\bibfield  {journal} {\bibinfo  {journal} {Rep. Prog. Phys.}\ }\textbf
  {\bibinfo {volume} {75}},\ \bibinfo {pages} {126001} (\bibinfo {year}
  {2012})}\BibitemShut {NoStop}%
\bibitem [{\citenamefont {Neri}\ \emph {et~al.}(2019)\citenamefont {Neri},
  \citenamefont {Rold{\'a}n}, \citenamefont {Pigolotti},\ and\ \citenamefont
  {J{\"u}licher}}]{Sneri2}%
  \BibitemOpen
  \bibfield  {author} {\bibinfo {author} {\bibfnamefont {I.}~\bibnamefont
  {Neri}}, \bibinfo {author} {\bibfnamefont {{\'E}.}~\bibnamefont
  {Rold{\'a}n}}, \bibinfo {author} {\bibfnamefont {S.}~\bibnamefont
  {Pigolotti}},\ and\ \bibinfo {author} {\bibfnamefont {F.}~\bibnamefont
  {J{\"u}licher}},\ }\bibfield  {title} {\emph {\bibinfo {title} {Integral
  fluctuation relations for entropy production at stopping times}},\
  }\href@noop {} {\bibfield  {journal} {\bibinfo  {journal} {J. Stat. Mech.}\
  }\textbf {\bibinfo {volume} {2019}},\ \bibinfo {pages} {104006} (\bibinfo
  {year} {2019})}\BibitemShut {NoStop}%
\bibitem [{\citenamefont {Manzano}\ \emph {et~al.}(2019)\citenamefont
  {Manzano}, \citenamefont {Fazio},\ and\ \citenamefont {Rold{\'a}n}}]{Sours}%
  \BibitemOpen
  \bibfield  {author} {\bibinfo {author} {\bibfnamefont {G.}~\bibnamefont
  {Manzano}}, \bibinfo {author} {\bibfnamefont {R.}~\bibnamefont {Fazio}},\
  and\ \bibinfo {author} {\bibfnamefont {{\'E}.}~\bibnamefont {Rold{\'a}n}},\
  }\bibfield  {title} {\emph {\bibinfo {title} {Quantum martingale theory and
  entropy production}},\ }\href@noop {} {\bibfield  {journal} {\bibinfo
  {journal} {Phys. Rev. Lett.}\ }\textbf {\bibinfo {volume} {122}},\ \bibinfo
  {pages} {220602} (\bibinfo {year} {2019})}\BibitemShut {NoStop}%
\end{thebibliography}
\end{document}